\documentclass[pre,showpacs,longbibliography,twocolumn]{revtex4-1}
\usepackage{graphicx}% Include figure files
\usepackage{color}% allows text in color with \textcolor{color}{text}
\usepackage{amsmath}
\usepackage{subfigure}
\usepackage{appendix}
\usepackage{soul}
\usepackage{bbm}
\usepackage[colorlinks,linkcolor=blue,citecolor=blue,urlcolor=blue,hyperindex,breaklinks]{hyperref}
\begin{document}
\title{Quantum thermal diode dominanted by pure classical correlation via three triangular-coupled qubits}
%\title{Pure classical correlation dominant quantum thermal diode via three triangular-coupled qubits}
\author{Yi-jia Yang, Yu-qiang Liu, and Chang-shui Yu}
\email{Electronic address: ycs@dlut.edu.cn}
\affiliation{School of Physics, Dalian University of Technology, Dalian 116024, China}
\date{\today}

\begin{abstract}
A quantum thermal diode is designed based on three pairwise coupled qubits, two connected to a common reservoir and the other to an independent reservoir. It is found that the internal couplings between qubits can enhance heat currents. If the two identical qubits uniformly couple with the common reservoir, the crossing dissipation will occur, leading to the initial-state-dependent steady state, which can be decomposed into the mixture of two particular steady states: the heat-conducting state generating maximum heat current and the heat-resisting state not transporting heat.  However, the rectification factor doesn't depend on the initial state.  In particular, we find that neither quantum entanglement nor quantum discord is present in the steady state, but the pure classical correlation shows a remarkably consistent behavior as the heat rectification factor, which reveals the vital role of classical correlation in the system.
\end{abstract}
%% Decoherence; open systems; quantum statistical methods, Quantum information, Quantum statistical mechanics, Thermodynamics

\maketitle
\section{introduction}
\label{sec1}
Thermal diode, similar to electrical diode \cite{PhysRevB.93.161410,li2012colloquium}, allows heat to transport only in a fixed direction and be blocked in the opposite direction  \cite{PhysRevLett.93.184301}. The asymmetric heat conductance in the solid-state nanosystem, consisting of highly conductive carbon and boron nitride nanotubes, is realized experimentally \cite{chang2006solid}. Recently, solid-state thermal diodes have been able to achieve considerable rectification ratio \cite{shrestha2020dual,luo2021heterogeneous,lee2021tunable}. A large number of asymmetric nanosystems such as monolayer graphene nanostructures \cite{wang2017experimental}, multiferroic helimagnet \cite{hirokane2020nonreciprocal}, have also been studied in thermal rectification.

Quantum thermal diode\cite{PhysRevLett.93.184301,werlang2014optimal,PhysRevE.95.022128,PhysRevB.98.035414,wang2019thermal,PhysRevB.99.035129,PhysRevE.99.042121,PhysRevE.104.054137,PhysRevE.103.052130,mojaveri2021maximal,PhysRevResearch.2.033285} has attracted increasing interest in quantum thermodynamics \cite{landsberg1956foundations,parrondo2015thermodynamics,millen2016perspective,strasberg2013thermodynamics,kosloff2013quantum,vinjanampathy2016quantum}, which has been studied extensively \cite{PhysRevLett.94.034301,cao2021quantum,ThreeLevelMasers,RevModPhys.84.1045,PhysRevE.94.042135,PhysRevLett.93.184301,seif2018thermal,ronzani2018tunable,martinez2015rectification,yang_2022} and provides an important platform to understand the quantum features in microscopic thermodynamic systems \cite{wang2002experimental,levy2014local,ronzani2018tunable,maillet2020electric,hewgill2021quantum}.
In particular, various microscopic thermal devices have been proposed, such as quantum Otto engine \cite{PhysRevLett.112.150602,PhysRevE.79.041113,PhysRevE.94.022141}, quantum thermometer \cite{hofer2017quantum,yang2019thermal,PhysRevLett.128.040502}, thermal memory \cite{PhysRevLett.101.267203}, quantum refrigerator \cite{PhysRevE.85.061126,PhysRevLett.110.256801,PhysRevE.90.052142,yu2019quantum}, quantum transistor \cite{joulain2016quantum,PhysRevE.94.042135,guo2018quantum,2019Multifunctional,PhysRevB.101.184510,PhysRevA.103.052613,PhysRevB.99.035129,PhysRevB.102.125405,wijesekara2020optically,PhysRevResearch.2.033285,wijesekara2021darlington,mandarino2021thermal,PhysRevE.106.024110}, quantum switch \cite{karimi2017coupled,farsani2019quantum,goswami2018indefinite}, and so on.
Unidirectional heat transfer is the key performance of a quantum thermal diode, which is characterized by the rectification factor. How to improve the rectification factor is one of the most important motivations in the design of various thermal diodes. The usual understanding of unidirectional heat transfer is attributed to the various asymmetries of the system.
For example, the diode system based on coupled two-level qubits which have different natural frequencies \cite{PhysRevE.95.022128}, or asymmetrical coupling such as $H_{SI}=\Delta\sigma_L^z\sigma_R^x$ \cite{PhysRevE.99.042121} with $\sigma^k$ denoting pseudo-spin operator of two-level atoms. The asymmetry can also come from the transition spectra induced by different environments \cite{PhysRevResearch.2.033285}.
It has recently been shown that common heat reservoirs can enhance the performance of quantum thermal devices and sometimes provide additional crossing dissipation channels \cite{PhysRevA.81.012105,PhysRevA.83.052110,PhysRevA.85.062323,PhysRevA.97.052309,hu2018steady,Cattaneo_2019,wang2019steady,wang2022nonequilibrium,mojaveri2021maximal}. This could also be significant for the design of a quantum thermal diode.
In addition, as a particular thermodynamical phenomenon, unidirectional heat transfer including many other quantum thermodynamical behaviors has not been well understood from the perspective of the features of quantum states  \cite{PhysRevE.79.041113,PhysRevE.80.061129,PhysRevE.104.054137,PhysRevB.98.035414,yang_2022}. It remains an open question of what feature of a quantum state is closely related to such thermodynamical phenomena. 
  
\begin{figure}
\begin{center}
\includegraphics[width=0.8\columnwidth]{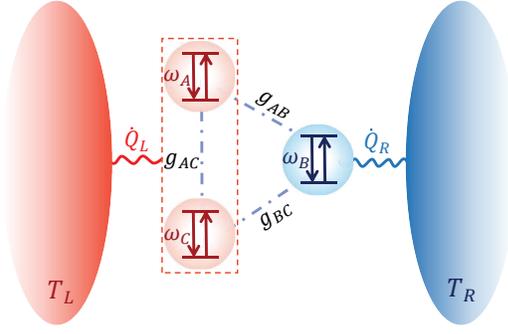}
\caption{Sketch of the quantum thermal diode. $\omega_\mu$ denotes the natural frequency of qubit $\mu$ and $g_{\mu\nu}$ with $\mu\neq \nu$ denotes the coupling strength between the qubits $\mu$ and $\nu$, $\mu, \nu=A,B,$ and $C$. $T_\alpha$ is the temperature of the heat reservoir $B_\alpha$, and $\dot{Q}_\alpha$ is the heat current from $B_\alpha$ to $S$ for $\alpha = L$ and $R$.}\label{model}
\end{center}
\end{figure}

This paper studies the heat transport and the thermal diode consisting of three pairwise coupled qubits interacting with two heat reservoirs of different temperatures \cite{joulain2016quantum,wijesekara2020optically,PhysRevA.103.052613}. Our system is sketched in Fig. \ref{model} where qubits A and C are commonly coupled to the left reservoir, and qubit B is in contact with the right reservoir. We find that the identical qubits A and C uniformly coupling with qubit B can produce crossing dissipations, leading to initial-state-dependent steady states. Moreover, the steady state can also be decomposed into the mixture of two particular steady states: the heat-conducting state generating maximum heat current and the heat-resisting state not transporting heat. It provides the potential to control the steady state by the initial state. We find that both the internal couplings of the qubits and the crossing dissipations can enhance the heat current through the left and right reservoirs, and the inverse heat currents corresponding to the crossing dissipations play a versatile role. In particular, we find the current model can act as a good quantum thermal diode, and the crossing dissipations aren't beneficial to the rectification effect. Importantly, we find that in the system, there's no quantum correlation (neither quantum entanglement nor quantum discord) between the system's left (AC) and right (B) parts; however, the classical correlation is closely related to the rectification effect.

This paper is organized as follows. We introduce our model and present the dynamics in Sec. \ref{sec2}. The steady states of the system and the heat currents between two reservoirs are presented in Sec. \ref{sec3}, and a quantum thermal diode is designed in Sec. \ref{sec4}. In Sec. \ref{sec5}, the relationship between the steady-state mutual information and rectification is studied. The conclusion and discussion are in Sec. \ref{sec6}. Some lengthy expressions are provided in Appendices, respectively.

\section{The Model and the dynamics}
\label{sec2}
Our model includes three qubits connected to two heat reservoirs, two commonly in contact with a common heat reservoir (CHR), and the remaining one to another independent heat reservoir (IHR), which is shown in Fig. \ref{model}.  The total Hamiltonian of the open system can be expressed as $H=H_S+H_E+H_{SE}$, where $(\hbar=\kappa_B=1)$ 
\begin{align}
H_S=\sum_\mu\frac{\omega_\mu}{2}\sigma^z_\mu+\sum_{\mu,\nu}\frac{g_{\mu\nu}}{2}\sigma^z_\mu\sigma^z_\nu,\quad\mu,\nu=A,B,C,\label{HS}
\end{align}
denotes the Hamiltonian of the system, $\omega_\mu$ represents the natural frequency of qubit $\mu$ and $g_{\mu\nu}$ is the coupling strength between qubits $\mu$ and $\nu$. Such an interaction mechanism has also been employed in some thermal devices as quantum thermal transistors \cite{joulain2016quantum,PhysRevA.103.052613}, and optically controlled thermal gate \cite{wijesekara2020optically}. $H_E$, the Hamiltonian of the reservoirs, reads 
\begin{equation}
H_E=\sum_\alpha\sum_k\omega_{\alpha k}a^\dagger_{\alpha k}a_{\alpha k},\quad\alpha=L,R,
\end{equation}
where $\omega_{\alpha k}$ is the frequency of the $k th$ mode of the heat reservoir $B_\alpha$, $a^\dagger_{\alpha k}$ (or $a_{\alpha k}$) denotes the corresponding creation (or annihilation) operator. 
The dipole interaction between the system and reservoir is given by
\begin{align}
\nonumber
H_{SE}&=\sum_k[(g_{Ak}\sigma^x_A+g_{Ck}\sigma^x_C)(a^\dagger_{Lk}+a_{Lk})\\
&+g_{Bk}\sigma^x_B(a^\dagger_{Rk}+a_{Rk})]\label{HSE},
\end{align}
where $g_{\mu k}$ is the coupling strength, characterizing the coupling capability between qubit $\mu$ and the $k th$ mode of the corresponding reservoir, and the operators $\sigma^x_\mu$ and $\sigma^z_\mu$ are the Pauli matrix with $\sigma^x_\mu=\left(
\begin{matrix}
0&1\\
1&0
\end{matrix}
\right)$ and $\sigma^z_\mu=\left(
\begin{matrix}
1&0\\
0&-1
\end{matrix}
\right)$.
It is obvious that $H_S$ has the diagonal form as $H_{S}=\sum_{i=1}^8\lambda_i\vert i\rangle\langle i\vert$, where the eigenstates are $\vert1\rangle=\vert+++\rangle$, $\vert2\rangle=\vert++-\rangle$, $\vert3\rangle=\vert+-+\rangle$, $\vert4\rangle=\vert+--\rangle$, $\vert5\rangle=\vert-++\rangle$, $\vert6\rangle=\vert-+-\rangle$, $\vert7\rangle=\vert--+\rangle$, and $\vert8\rangle=\vert---\rangle$, with $\vert \pm\pm\pm\rangle=\vert \pm\rangle_A\otimes\vert \pm\rangle_B\otimes\vert \pm\rangle_C$ and $\vert +\rangle_\mu$ (or $\vert -\rangle_\mu$) corresponds to the up-spin (or down-spin) eigenstate of the $\mu th$ qubit, respectively, and the eigenvalues $\lambda_i$ are shown in Appendix \ref{AppendixA}. 

\begin{figure}
\begin{center}
		\includegraphics[width=8cm]{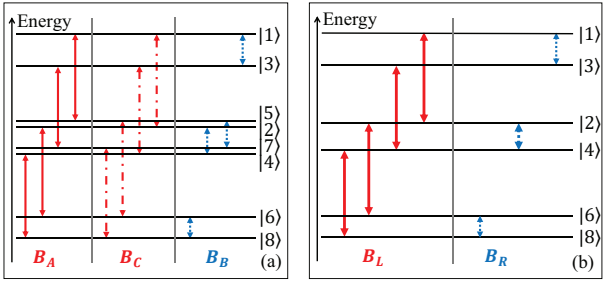}
\caption{Energy levels and transitions with $\omega_A=\omega_C=\omega$ and $g_{AB}=g_{BC}=g$. The Red solid and dot-dashed lines indicate the qubits A and C transitions induced by reservoir $L$, respectively, and the blue arrows indicate the qubit B transitions induced by reservoir $R$. (b) Variational transition diagram with degenerate levels reduced to a single one due to no transition between degenerate energy levels. }
\label{level}
\end{center}
\end{figure}

The crossing dissipation is inevitable, when the common environment induces two subsystems connected to it to undergo transitions of the same frequency \cite{PhysRevA.83.052110,hu2018steady,wang2019steady,Cattaneo_2019,mojaveri2021maximal,yang_2022}. By analyzing the eigenenergy levels of the triangle-coupled system Eq. (\ref{Eqlevel}), we find that there is crossing dissipation between the heat reservoir $B_L$ and the system when $\omega_A=\omega_C=\omega$ and $g_{AB}=g_{BC}=g$, called \textit{crossing condition}, are satisfied.
Following the standard process \cite{breuer2002theory}, one can obtain the Born-Markov-secular master equation for the system's reduced density matrix $\rho(t)$ as
\begin{equation}
\dot{\rho}(t)=-i[H_S,\rho(t)]+\sum_{\mu=A,B,C}\mathcal{L}_\mu[\rho(t)]+\mathcal{L}_{AC}[\rho(t)],\label{MEq}
\end{equation}
the superoperator $\mathcal{L}_\mu[\rho(t)]$ reads
\begin{align}
\nonumber
\mathcal{L}_\mu[\rho(t)]=&\sum_{(i,j)} J_\mu(-\omega_{ij}^\mu)[2V^\mu_{ij}\rho(t) {V^\mu_{ij}}^\dagger-\{{V^\mu_{ij}}^\dagger V^\mu_{ij},\rho(t)\}]\\
+&J_\mu(+\omega_{ij}^\mu)[2{V^\mu_{ij}}^\dagger\rho(t) V^\mu_{ij}-\{V^\mu_{ij}{V^\mu_{ij}}^\dagger,\rho(t)\}],\label{dissipation}
\end{align}
and $\mathcal{L}_{AC}[\rho(t)]$ denotes the potential crossing dissipator given by
\begin{align}
\nonumber
\mathcal{L}_{AC}[\rho(t)]=&\sum_{(i,j)} J_{AC}(-\omega_{ij}^{AC})[2V_{ij}^A\rho(t) {V_{ij}^C}^\dagger-\{{V_{ij}^C}^\dagger V_{ij}^A,\rho(t)\}\\
\nonumber
+&2V_{ij}^C\rho(t) {V_{ij}^A}^\dagger-\{{V_{ij}^A}^\dagger V_{ij}^C,\rho(t)\}]\\
\nonumber
+&J_{AC}(+\omega_{ij}^{AC})[2{V_{ij}^A}^\dagger\rho(t) V_{ij}^C-\{V_{ij}^C{V_{ij}^A}^\dagger,\rho(t)\}\\
+&2{V_{ij}^C}^\dagger\rho(t) V_{ij}^A-\{V_{ij}^A{V_{ij}^C}^\dagger,\rho(t)\}],\label{dissipationAandC}
\end{align}
where $J_{AC}(\pm\omega_{ij}^{AC})=\sqrt{J_A(\pm\omega_{ij}^A)J_C(\pm\omega_{ij}^C)}$, $\{\cdot,\cdot\}$ denotes the anticommutation relation, $V_{ij}^\mu=\vert j\rangle\langle i\vert$ and $\omega_{ij}^\mu=\lambda_i-\lambda_j$
represent the transition operator (or eigen-operator) of the qubit $\mu$ and the eigenfrequency, and $\sum_{(i,j)}$ means summation over all dissipators of the qubit $\mu$. A simple calculation can show that for each qubit, only four eigen-operators, i.e., $V_{15}^A$, $V_{26}^A$, $V_{37}^A$, $V_{48}^A$ for qubit $A$, $V_{13}^B$, $V_{24}^B$, $V_{57}^B$, $V_{68}^B$ for qubit $B$, and $V_{12}^C$, $V_{34}^C$, $V_{56}^C$, $V_{78}^C$ for qubit $C$ can be present. A diagrammatic sketch of these transitions is given in Fig. \ref{level}\hyperref[level]{(a)}.
$J_\mu(\pm\omega^\mu_{ij})=\kappa_\mu(\omega_{ij}^\mu)[\pm \bar{n}_\mu(\pm\omega_{ij}^\mu)]$ denotes the spectral density, where $\kappa_\mu(\omega^\mu_{ij})=\kappa$ is supposed to be the flat spectrum for simplicity in this work. $\bar{n}_\mu(\omega_{ij}^\mu)=(e^{{\omega_{ij}^\mu}/{T_\mu}}-1)^{-1}$ is the average photon number with $T_\mu$ denoting the temperature of the heat reservoir connected to the qubit $\mu$.
In addition, the dissipators for the left and right heat reservoirs can be written as
\begin{align}
\nonumber
\mathcal{L}_L[\rho(t)]&=\mathcal{L}_A[\rho(t)]+\mathcal{L}_C[\rho(t)]+\mathcal{L}_{AC}[\rho(t)],\\
\mathcal{L}_R[\rho(t)]&=\mathcal{L}_B[\rho(t)].\label{RLres}
\end{align}
Here, we must emphasize that $\mathcal{L}_{AC}[\rho(t)]$ in Eq. (\ref{MEq}) occurs only under the crossing condition, where the energy levels $\vert 2\rangle$ and $\vert 5\rangle$, $\vert 4\rangle$ and $\vert 7\rangle$ are degenerate. All transitions induced by the two reservoirs are shown in Fig. \ref{level}\hyperref[level]{(b)}.
If the energy levels are not degenerate, $\mathcal{L}_{AC}[\rho]=0$, which means that qubits $A$ and $C$ are in contact with two identical IHRs. In the following, we use the superscripts $I$ to indicate IHR, and $C$, correspondingly, to denote the presence of crossing dissipations, i.e., the genuine CHR.
In order to distinguish between IHR and CHR, when the crossing dissipation exists, we let $J_L(\pm\omega^L_{ij})=J_\mu(\pm\omega^\mu_{ij})$ for $\mu=A,C$ and $J_R(\pm\omega^R_{ij})=J_B(\pm\omega^B_{ij})$ to highlight the left (L) and right (R) heat reservoirs.

\section{Steady state and heat currents}
\label{sec3}
\subsection{Steady state}
The evolution of the density matrix can be divided into two independent subspaces. All the relevant matrix elements in one subspace will vanish for the steady-state case. In the other subspace, non-zero matrix elements to be obtained dominate the steady-state contribution, the relevant evolution equations are given as follows,
\begin{align}
\nonumber
\dot{\rho}_{11}=&-\Gamma^A_{15}-\Gamma^B_{13}-\Gamma^C_{12}+2\sqrt{A_{15}^+C_{12}^+}(\rho_{25}+\rho_{52}),\\
\nonumber
\dot{\rho}_{33}=&-\Gamma^A_{37}+\Gamma^B_{13}-\Gamma^C_{34}+2\sqrt{A_{37}^+C_{34}^+}(\rho_{47}+\rho_{74}),\\
\nonumber
\dot{\rho}_{66}=&+\Gamma^A_{26}-\Gamma^B_{68}+\Gamma^C_{56}+2\sqrt{A_{26}^-C_{56}^-}(\rho_{25}+\rho_{52}),\\
\nonumber
\dot{\rho}_{88}=&+\Gamma^A_{48}+\Gamma^B_{68}+\Gamma^C_{78}+2\sqrt{A_{48}^-C_{78}^-}(\rho_{47}+\rho_{74}),\\
\nonumber
\dot{\rho}_{22}=&-\Gamma^A_{26}-\Gamma^B_{24}+\Gamma^C_{12}\\
\nonumber
-&(\sqrt{A_{15}^+C_{12}^+}+\sqrt{A_{26}^-C_{56}^-})(\rho_{25}+\rho_{52}),\\
\nonumber
\dot{\rho}_{55}=&+\Gamma^A_{15}-\Gamma^B_{57}-\Gamma^C_{56}\\
\nonumber
-&(\sqrt{A_{15}^+C_{12}^+}+\sqrt{A_{26}^-C_{56}^-})(\rho_{25}+\rho_{52}),\\
\nonumber
\dot{\rho}_{44}=&-\Gamma^A_{48}+\Gamma^B_{24}+\Gamma^C_{34}\\
\nonumber
-&(\sqrt{A_{37}^+C_{34}^+}+\sqrt{A_{48}^-C_{78}^-})(\rho_{47}+\rho_{74}),\\
\nonumber
\dot{\rho}_{77}=&+\Gamma^A_{37}+\Gamma^B_{57}-\Gamma^C_{78}\\
\nonumber
-&(\sqrt{A_{37}^+C_{34}^+}+\sqrt{A_{48}^-C_{78}^-})(\rho_{47}+\rho_{74}),\\
\nonumber
\dot{\rho}_{25}=&\Gamma^{AC}_{1,52}-\Gamma^{AC}_{25,6}+2\sqrt{B_{24}^+B_{57}^+}\rho_{47}\\
\nonumber
-&(A_{15}^++A_{26}^-+B_{24}^-+B_{57}^-+C_{12}^++C_{56}^-)\rho_{25},\\
\nonumber
\dot{\rho}_{47}=&\Gamma^{AC}_{3,74}-\Gamma^{AC}_{47,8}+2\sqrt{B_{24}^-B_{57}^-}\rho_{25}\\
%\nonumber
-&(A_{37}^++A_{48}^-+B_{24}^++B_{57}^++C_{34}^++C_{78}^-)\rho_{47},\label{xcv}
\end{align}
$\dot{\rho}_{52}=\dot{\rho}_{25}^*$ and $\dot{\rho}_{74}=\dot{\rho}_{47}^*$, where $\Gamma_{ij}^\mu=2[J_\mu(-\omega_{ij}^\mu)\rho_{ii}-J_\mu(+\omega_{ij}^\mu)\rho_{jj}]$ denotes the net dissipation rate from $\vert i\rangle$ to $\vert j\rangle$ corresponding to the qubit $\mu$ and $\rho_{ii}$ indicates the population of $\vert i\rangle$.
$\mu_{ij}^\pm$ is a simplified representation of the spectral density $J_\mu(\pm\omega_{ij}^\mu)$.
The net dissipation rate induced by CHR $\Gamma_{ijk}^{AC}$ are non-vanished and can be expressed as 
\begin{align}
\Gamma^{AC}_{1,25}&=2\sqrt{A_{15}^-C_{12}^-}\rho_{11}-\sqrt{A_{15}^+C_{12}^+}(\rho_{22}+\rho_{55}),\notag\\
\Gamma^{AC}_{3,47}&=2\sqrt{A_{37}^-C_{34}^-}\rho_{33}-\sqrt{A_{37}^+C_{34}^+}(\rho_{44}+\rho_{77}),\notag\\
\Gamma^{AC}_{25,6}&=\sqrt{A_{26}^-C_{56}^-}(\rho_{22}+\rho_{55})-2\sqrt{A_{26}^+C_{56}^+}\rho_{66},\notag\\
\Gamma^{AC}_{47,8}&=\sqrt{A_{48}^-C_{78}^-}(\rho_{44}+\rho_{77})-2\sqrt{A_{48}^+C_{78}^+}\rho_{88}.
\end{align}

If the conditions $\omega_A=\omega_C=\omega$ and $g_{AB}=g_{BC}=g$ aren't satisfied, there is no crossing dissipation. Thus the above evolution equations will be simplified, by setting $\dot{\rho}_{52}=\dot{\rho}_{74}=0$, as $\vert\dot{\rho}^I\rangle=\mathcal{M}^I\vert\rho^I\rangle$, where $\vert\rho^I\rangle=[\rho_{11}^I,\rho_{22}^I,\rho_{33}^I,\rho_{44}^I,\rho_{55}^I,\rho_{66}^I,\rho_{77}^I,\rho_{88}^I]^T$ is the column vector of non-zero diagonal entries of the density matrix and the coefficient matrix $\mathcal{M}^I=\mathcal{M}_A^I+\mathcal{M}_B^I+\mathcal{M}_C^I$, which is concretely expressed in Appendix \ref{AppendixB}.
The equations can be analytically solved, and the steady state can be written as $\rho^{I,S}=\sum_{i=1}^8\rho_{ii}^I\vert i\rangle\langle i\vert$. However, the explicit form is too tedious and we don't present it here.

The crossing dissipations will be present under the crossing condition, which means that the CHR will take effect.
Through the dynamics of the system Eq. (\ref{xcv}), it can be obtained that the degenerate energy levels have the same population at steady state, that is, $\rho^C_{22}=\rho^C_{55}$ and $\rho^C_{44}=\rho^C_{77}$.
Similarly, the density matrix for the CHR can be written $\vert\dot{\rho}^C\rangle=\mathcal{M}^C\vert\rho^C\rangle$ with $\vert\rho^C\rangle=[\rho^C_{11},\rho^C_{22},\rho^C_{33},\rho^C_{44},\rho^C_{66},\rho^C_{88},\rho^C_{25},\rho^C_{47}]^T$, and $\mathcal{M}^C=\mathcal{M}_L^C+\mathcal{M}_R^C$, which is shown in Appendix \ref{AppendixB} for their complex expressions.
One can find that the rank of $\mathcal{M}^C$ is $6$, but the dimension of $\mathcal{M}^C$ is $8$. So the steady states are not unique and depend on the initial state. 
In other words, the steady state of the system can always be written as the mixture of two particular steady states as
\begin{align}
\rho^{C,S}=(1-p)\rho^{C,S}_1+p\rho^{C,S}_2,\label{steady0}
\end{align}
where $\rho_1^{C,S}=\sum_{i=1}^2\tilde{\rho}_{ii}^S\vert \tilde{i}\rangle\langle\tilde{i}\vert$ and $\rho_2^{C,S}=\sum_{i=3}^8\tilde{\rho}_{ii}^S\vert \tilde{i}\rangle\langle\tilde{i}\vert$, the specific expression of $\tilde{\rho}_{ii}^S$ is shown in Appendix \ref{AppendixB}, and $\{\vert \tilde{i}\rangle\}$ are alternative eigenstates of the system defined as
\begin{align}
\nonumber
\vert\tilde{1}\rangle&=\frac{1}{\sqrt{2}}(\vert5\rangle-\vert2\rangle),&\quad\vert\tilde{5}\rangle&=\frac{1}{\sqrt{2}}(\vert2\rangle+\vert5\rangle),\\
\nonumber
\vert\tilde{2}\rangle&=\frac{1}{\sqrt{2}}(\vert7\rangle-\vert4\rangle),&\quad\vert\tilde{6}\rangle&=\frac{1}{\sqrt{2}}(\vert4\rangle+\vert7\rangle),\\
\vert\tilde{3}\rangle&=\vert1\rangle,\  \vert\tilde{4}\rangle=\vert3\rangle,&\ \vert\tilde{7}\rangle&=\vert6\rangle,\ \vert\tilde{8}\rangle=\vert8\rangle.\label{newbasis}
\end{align}
In particular, the initial state can also be written as $\rho^0=(1-p)S_1+pS_2$ with $S_1=\sum_{i,j=1}^2a_{ij}\vert \tilde{i}\rangle\langle\tilde{j}\vert$ and $S_2=\sum_{i,j=3}^8b_{ij}\vert \tilde{i}\rangle\langle\tilde{j}\vert$ representing density matrices in the corresponding subspace $\{\left\vert \tilde{i}\right\rangle\}_{i=1,2}$ and $\{\left\vert \tilde{i}\right\rangle\}_{i=3,\cdots,8}$, and $p=\sum_{i=3}^8\left\langle \tilde{i}\right\vert \rho^0\left\vert\tilde{i}\right\rangle$ quantifies the fraction. We have obtained the steady-state density matrices in both CHR and IHR cases to study our system's thermodynamic behaviors. In particular, it allows controlling the steady state of our demand by adjusting the initial state.

\subsection{Steady-state heat current}
\begin{figure}
\begin{center}
		\includegraphics[width=8cm]{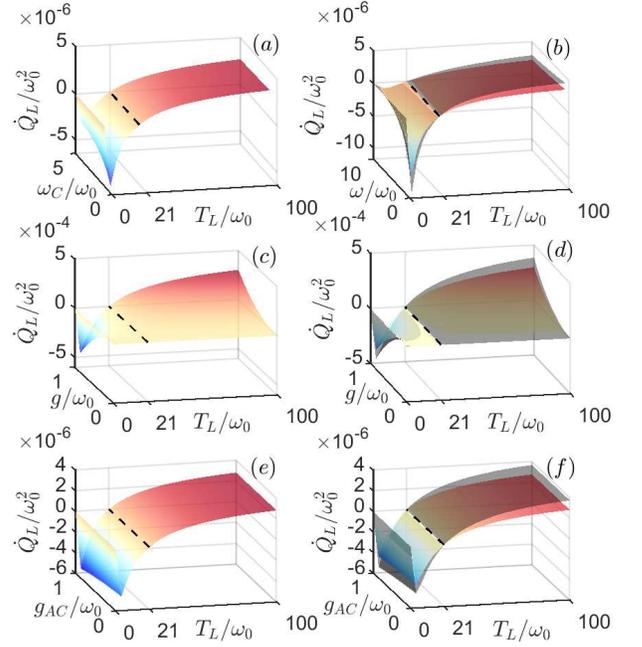}
\caption{Steady-state heat current $\dot{Q}_L$ with temperature $T_L$. $\dot{Q}_L$ and $\dot{Q}_R$ are of the same magnitude and opposite directions. The black dashed lines indicate $\dot{Q}_L=\dot{Q}_R=0$ for $T_L=T_R$. The left three figures correspond to $\omega_A=3\omega_0$ and $\omega_C=2\omega_0$. The right three figures correspond to $\omega_A=\omega_C=\omega$, where the shaded (outside) and colored (inside) surfaces represent the steady-state heat current with and without crossing dissipation. Here, $\omega_0=1$, $\omega=3\omega_0$, $\omega_B=5\omega_0$, $g_{AB}=g_{BC}=g_{AC}=0.1\omega_0$, $\kappa=0.001\omega_0$, $T_R=21\omega_0$, and $p=1$ in (b), (d), and (f).}
\label{Q}
\end{center}
\end{figure}
The heat current between the heat reservoir and the contact qubit $\mu$ is defined as
\begin{align}
\dot{Q}_\mu=\mathrm{Tr}\{H_S\mathcal{L}_\mu[\rho(t)]\},\quad\mu=A,B,C.\label{Def_QABC}
\end{align}
{$\dot{Q}_\mu<0$ means that heat flows from the qubit $\mu$ into its heat reservoir}, and oppositely, it indicates heat flows out of the reservoir to our system of interest.
The heat current for the reservoir $\alpha$ can be given by \begin{align}
\dot{Q}_\alpha=\mathrm{Tr}\{H_S\mathcal{L}_\alpha[\rho(t)]\},\quad\alpha=L,R,\label{Def_Q}
\end{align}
with $\mathcal{L}_\alpha[\rho(t)]$ defined in Eq. (\ref{RLres}). $\dot{Q}_\alpha$ describes the net heat current into or out of the reservoir $B_\alpha$.

In the case of IHR, the steady-state heat current $\dot{Q}_\mu\vert_{\rho^{I,S}}$ can be obtained by substituting state $\rho^{I,S}$ into Eq. (\ref{Def_QABC}), which is specifically shown in Appendix \ref{AppendixC}.
It is obvious that $\dot{Q}_L\vert_{\rho^{I,S}}=\dot{Q}_A\vert_{\rho^{I,S}}+\dot{Q}_C\vert_{\rho^{I,S}}=-\dot{Q}_B\vert_{\rho^{I,S}}=-\dot{Q}_R\vert_{\rho^{I,S}}$ corresponds to the law of conservation of energy.

Similarly, in the case of CHR, one will find that the heat currents for the steady state $\rho_1^{C,S}$ are
\begin{align}
\dot{Q}_L\vert_{\rho_1^{C,S}}=\dot{Q}_R\vert_{\rho_1^{C,S}}=0,\label{Q=0}
\end{align}
which means $\rho_1^{C,S}$ blocks heat transfer between the two reservoirs and is a heat-resisting state. 
The heat currents for the steady state $\rho_2^{C,S}$, i.e., $\dot{Q}_\alpha\vert_{\rho_2^{C,S}}$, can also be easily obtained, which are given in Appendix \ref{AppendixC}.
Since the steady state can be written as the mixture of the two states $\rho_1^{C,S}$ and $\rho_2^{C,S}$ as Eq. (\ref{steady0}), the steady-state heat currents can also be the mixture of the heat currents of the two particular states. That is, the steady-state heat current can be expressed as
\begin{align}
\dot{Q}_\alpha\vert_{\rho^{C,S}}=p\dot{Q}_\alpha\vert_{\rho^{C,S}_2}.\label{steady_Qp}
\end{align} 
One can see that the heat current depends on the fraction $p$. In this sense, one can immediately find that the heat current  $\dot{Q}_\alpha\vert_{\rho_2^{C,S}}$ for the state $\rho_2^{C,S}$ corresponds to the maximal potential steady-state heat current. Therefore, the steady-state heat current can be effectively controlled by the fraction of the initial states.

To reveal the effect of the system's structure of our model on the heat currents, we plot heat current versus the temperature with different internal parameters in Fig. \ref{Q}. In Fig. \ref{Q}\hyperref[Q]{(a)} we fix the frequencies $\omega_A=3\omega_0$ and $\omega_B=5\omega_0$ to show how the heat currents depend on the frequency $\omega_C$. It can be easily found that the heat current is reduced with increasing the frequency $\omega_C$. Similarly, one can find that the internal couplings $g$ and $g_{AC}$ can enhance the heat currents, which are shown in Figs. \ref{Q}\hyperref[Q]{(c)} and \ref{Q}\hyperref[Q]{(e)}, respectively. But comparing the two figures indicates that the enhancement effect of the coupling $g_{AC}$ between $A$ and $C$ seems weaker than that of $g$. It can be well understood because $g$ characterizes the contact intensity of the system's left and right parts, but $g_{AC}$ characterizes the coupling between the two transport channels $A \leftrightarrow B$ and $C \leftrightarrow B$.
Of course, it can be trivially understood that $g=0$ will show zero heat current due to two separate left and right reservoirs, which case has been verified through an analytic process in Appendix \ref{AppendixB}. 
As mentioned previously, if $\omega_A=\omega_C=\omega$ and $g_{AB}=g_{BC}=g$, the crossing dissipations will occur and will play its role. To illustrate the effect of CHR, we plot the heat currents with CHR and $p=1$ by the shaded surfaces in Figs. \ref{Q}\hyperref[Q]{(b)}, \ref{Q}\hyperref[Q]{(d)}, and \ref{Q}\hyperref[Q]{(f)}. As a comparison, we also consider that two identical qubits $A$ and $C$ are in contact with a heat reservoir $L$, respectively. The colored surfaces in the figures also plot the corresponding heat currents. 
The CHR can be found to enhance the heat currents, and the enhancement effect becomes significant with the increase of heat currents.
\begin{figure}
\begin{center}
		\includegraphics[width=8cm]{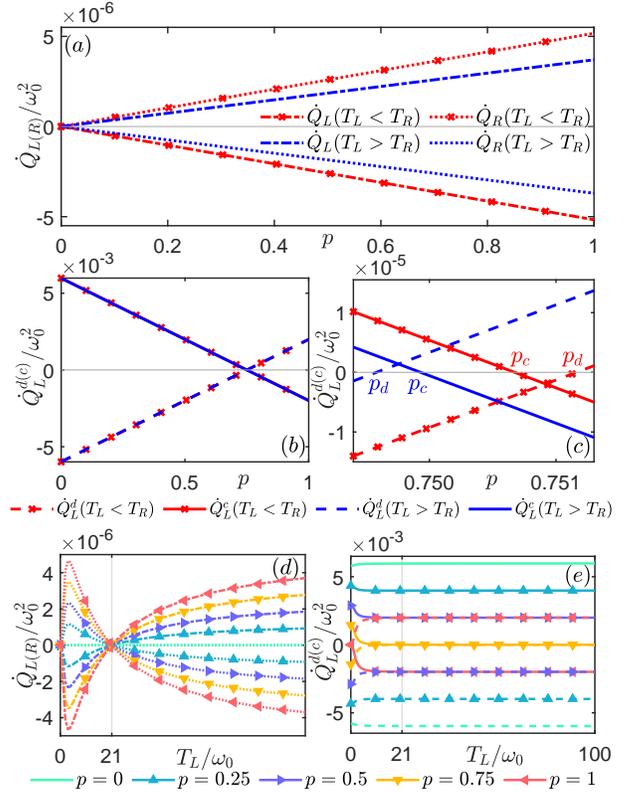}
\caption{Heat currents versus the fraction $p$ in (a-c) and the temperature $T_L$ in (d-e). Here $\dot{Q}_L^d+\dot{Q}_L^c=\dot{Q}_L$. (c) is the enlarged view of the region around the ''cross point'' in (b). The gray lines in (d) and (e) correspond to $T_L=T_R$. In (a) and (d), $\dot{Q}_\alpha, \alpha=L, R,$ with dotted-dashed line and dotted line represent the NHC between two reservoirs. In (b), (c), and (e), $\dot{Q}_L^d$ with dashed lines and $\dot{Q}_L^c$ with solid lines denote the DHC and CHC. $T_L>T_R$ means $T_L=100\omega_0$ and $T_R=21\omega_0$, and $T_L<T_R$ means $T_R=100\omega_0$ and $T_L=21\omega_0$. The other parameters are taken the same as Fig. \ref{Q}.}
\label{DHC_CHC}
\end{center}
\end{figure}

\subsection{Inverse heat current}
In the case of CHR, crossing dissipators will exist in the master equation. Usually, the different dissipators determine different heat transport channels and play different roles in the heat currents. To refine the effects of these dissipators, we can define the direct dissipation channel determined by $\mathcal{L}^d_L[\rho(t)]=\mathcal{L}_A[\rho(t)]+\mathcal{L}_C[\rho(t)]$ and the crossing dissipation channel governed by $\mathcal{L}^c_L[\rho(t)]=\mathcal{L}_{AC}[\rho(t)]$. Analogously to Eq. (\ref{Def_QABC}), we can also define the direct heat current (DHC) and the crossing heat current (CHC) corresponding to direct and crossing dissipation channels, respectively.

As mentioned previously, $\rho_1^{C,S}$ (or the steady state $p=0$) is a heat-resisting state that blocks the system's heat transfer. Since $\dot{Q}_L\vert_{\rho_1^{C,S}}=0$, considering DHC and CHC, one can also divide the vanishing heat current of $\rho_1^{C,S}$ into two parts of the same magnitude and opposite direction as $\dot{Q}^d_L\vert_{\rho_1^{C,S}}=-\dot{Q}^c_L\vert_{\rho_1^{C,S}}$, where the superscript $d$ and $c$ label DHC and CHC, respectively. Analogously, the heat current for the state $\rho_2^{C,S}$ can also be divided as $\dot{Q}_L\vert_{\rho_2^{C,S}}=\dot{Q}_L^d\vert_{\rho_2^{C,S}}+\dot{Q}_L^c\vert_{\rho_2^{C,S}}$. The expressions of the heat current in the two channels corresponding to the two sub-steady states are explicitly shown in Appendix \ref{AppendixC}.
Note that $\dot{Q}_L^c\vert_{\rho^{C,S}_2}\neq-\dot{Q}_L^d\vert_{\rho^{C,S}_2}$ is different from the state $\rho^{C,S}_1$. 
Accordingly, the total heat current $\dot{Q}_\alpha$ can also be divided in DHC and CHC as
\begin{align}
\dot{Q}_L^{d/c}=(1-p)\dot{Q}_L^{d/c}\vert_{\rho_1^{C,S}}+p\dot{Q}_L^{d/c}\vert_{\rho_2^{C,S}},\label{Qdc}
\end{align}
which is also proportional to the fraction $p$. 

DHC or CHC opposite to the direction of the net heat current (NHC) is referred to as the inverse heat current (IHC) \cite{wang2020inverse,yang_2022}, which could play some subtle role in the heat transfer. To show this, we set the left reservoir $B_L$ as the hot (and cold) terminal, respectively, and plot DHC, CHC, and NHC in Fig. \ref{DHC_CHC}. It is obvious in Fig. \ref{DHC_CHC}\hyperref[DHC_CHC]{(a)} that the NHC is conserved, that is, $\dot{Q}_L=-\dot{Q}_R$ for both $T_L>T_R$ and $T_R>T_L$, and the NHC linearly depends on the fraction $p$. But DHC and CHC compete for the dominant role played in the NHC, which can be seen from Fig. \ref{DHC_CHC}\hyperref[DHC_CHC]{(b)} and Fig. \ref{DHC_CHC}\hyperref[DHC_CHC]{(c)} that DHC and CHC have variant directions.
In particular, from Fig. \ref{DHC_CHC}\hyperref[DHC_CHC]{(c)} one can find that $\dot{Q}_L^d=0$ and $\dot{Q}_L^c=0$ at the points
\begin{align}
p_d=\frac{\dot{Q}_L^d\vert_{\rho^{C,S}_1}}{\dot{Q}_L^d\vert_{\rho^{C,S}_1}-\dot{Q}_L^d\vert_{\rho^{C,S}_2}},\quad p_c=\frac{\dot{Q}_L^d\vert_{\rho^{C,S}_1}}{\dot{Q}_L^d\vert_{\rho^{C,S}_1}+\dot{Q}_L^c\vert_{\rho^{C,S}_2}}.\label{pdc}
\end{align} 
$p_{d(c)}$ corresponds to the point of the vanishing heat current in the direct dissipation (or crossing dissipation) channel. Namely, at the point $p=p_d$, $\dot{Q}_L^d=0$ and $\dot{Q}_L^c=\dot{Q}_L$, and at the point $p=p_c$, $\dot{Q}_L^c=0$ and $\dot{Q}_L^d=\dot{Q}_L$. Thus one can easily see that the dissipative channels play different roles in NHC in different ranges of $p$.
Let's first focus on the case $T_L>T_R$ (blue lines). When $p<p_d$, CHC plays a dominant role in the NHC, but $\dot{Q}_L^d$ serves as the inverse heat current. When $p>p_c$, DHC plays a dominant role in NHC, and $\dot{Q}_L^c$ is the inverse heat current. However, when $p\in [p_d,p_c]$, $\dot{Q}_L^d>0$ and $\dot{Q}_L^c>0$ and there is no IHC between the system and the bath. In the case of $T_L<T_R$ (red lines), the phenomena are quite similar to the case of $T_L>T_R$. The differences are that NHCs change the directions due to exchanging the hot and the cold terminals as shown in Fig. \ref{DHC_CHC}\hyperref[DHC_CHC]{(a)}, and the points $p_{d}$ and $p_{c}$ exchange their relative positions, which would result in NHC being dominated by DHC and CHC in $p\in[0,p_c]$ and $p\in[p_d,1]$.
In Fig. \ref{DHC_CHC}\hyperref[DHC_CHC]{(d)}, one can see that the heat current always flows from the hot end to the cold end and the total heat current is zero due to the conservation of energy.
It is apparent that $T_L=T_R$ means the thermal equilibrium, so there is no heat current. It is interesting that, if either reservoir approaches $0$K, the NHC also vanishes. A detailed mathematical derivation is given by Eqs. (\ref{Qcd2TL}, \ref{Qcd2TR}).
Fig. \ref{DHC_CHC}\hyperref[DHC_CHC]{(e)} indicates that the heat currents between $B_L$ and the system in the two channels are not sensitive to the temperature. The robustness of $\dot{Q}_L^{d(c)}$ to $T_L$ increases with the decrease of the fraction $p$, which is quite clear at the range of $T_L\rightarrow 0$. Since NHC increases with $p$ increasing, $p=0$ means that NHC vanished, which is reasonable even if there is a temperature gradient between the two reservoirs.
\begin{figure}
\begin{center}
		\includegraphics[width=8cm]{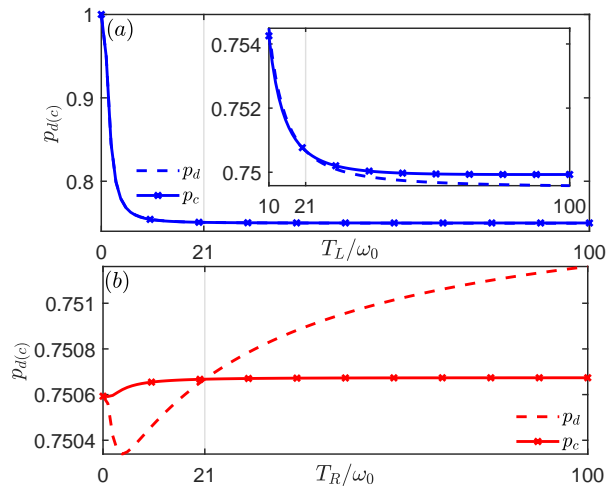}
\caption{$p_d$ and $p_c$ versus the temperature of one terminal. The inset in (a) is a local magnification around $T_L/\omega_0\in[10,100]$. In both figures, we set the fixed temperature $21\omega_0$ to the other terminal. The other parameters are taken the same as Fig. \ref{Q}.}
\label{pcpd_TL}
\end{center}
\end{figure}

Fig. \ref{pcpd_TL}\hyperref[pcpd_TL]{(a)} (or Fig. \ref{pcpd_TL}\hyperref[pcpd_TL]{(b)}) depicts $p_c$ and $p_d$ depending on the temperature $T_L$ (or $T_R$) with the other terminal fixed to be constant temperature $21\omega_0$. It is easy to see that $p_c$ and $p_d$ greatly depend on the temperature for $T_{L(R)}<10\omega_0$. However, $p_c$ is insensitive to the temperature for $T_{L(R)}>10\omega_0$. In addition, one can also find that $p_c$ and $p_d$ demonstrate different behaviors with the temperature when the temperature of the two terminals exchanged (corresponding to Fig. \ref{pcpd_TL}\hyperref[pcpd_TL]{(a)} and Fig. \ref{pcpd_TL}\hyperref[pcpd_TL]{(b)}). In fact, $p_c$ and $p_d$ also depend on the other parameters of the system, such as atomic natural frequency $\omega(\omega_B)$, interatomic coupling strength $g(g_{AC})$, which are given in Appendix \ref{AppendixD}.

\section{Quantum thermal diode} 
\label{sec4}
Electric diodes, important electric devices, allow the electric current to conduct only in one direction. Similarly, quantum thermal diodes can only allow unidirectional transport of the heat current between two reservoirs. The rectification performance can be characterized by the rectification (asymmetry) factor defined as 
\begin{align}
\mathcal{R}_\alpha=\left|\frac{\left|\dot{Q}_\alpha^f\right|-\left|\dot{Q}_\alpha^r\right|}{\mathrm{Max}[\dot{Q}_\alpha^f,\dot{Q}_\alpha^r]}\right|,\label{defination_R}
\end{align}
where $\dot{Q}_\alpha^{f}$ denotes the forward heat current defined by $\dot{Q}_\alpha$ for given $T_L$ and $T_R$, and $\dot{Q}_\alpha^{r}$ denotes the reverse heat current defined by $\dot{Q}_\alpha$ with exchanging the $T_L$ and $T_R$ for $\dot{Q}_\alpha^{f}$. The energy conservation implies $\mathcal{R}_L=\mathcal{R}_R=\mathcal{R}$. $\mathcal{R}\in[0,1]$, where $\mathcal{R}=1$ means the perfect diode and $\mathcal{R}=0$ means the vanishing rectification. 

\begin{figure}
\begin{center}
		\includegraphics[width=8cm]{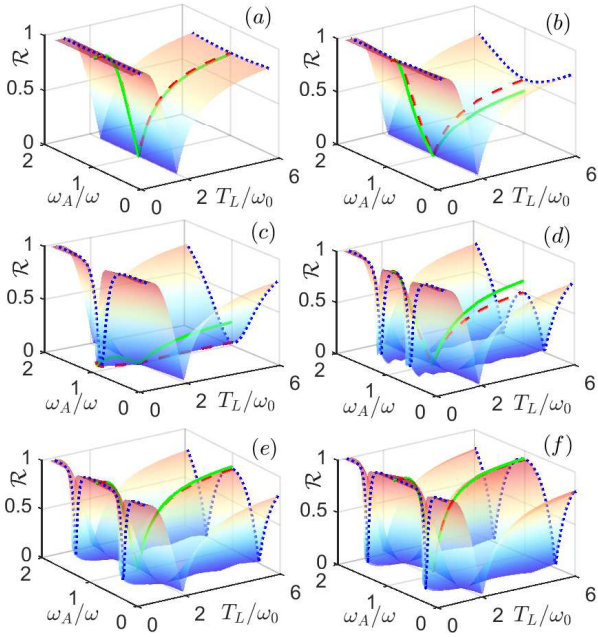}
\caption{Rectification factor $\mathcal{R}$ versus $\omega_A$ and $T_L$. The blue dotted lines correspond to $T_L=0.4\omega_0$ and $6\omega_0$, the red dashed lines correspond to $\omega_A=\omega_C=\omega$, and the green solid lines correspond to the case in the presence of the crossing dissipation $\mathcal{L}_{AC}(\rho)$. Here, $\omega_C=2\omega-\omega_A$, $\omega_0=1$, $\kappa=0.001\omega_0$, $p=1$, $T_R=2\omega_0$, $\omega_B=5\omega_0$, $g_{AB}=g_{BC}=g_{AC}=0.1\omega_0$, and $\omega=\omega_0,3\omega_0,5\omega_0,7\omega_0,10\omega_0,12\omega_0$ in (a-f).}
\label{R_3D}
\end{center}
\end{figure}

In general, the diode function comes from the asymmetry of the quantum system, such as the differences of the coupling $g_{\mu\nu}$ between qubits, of the natural frequencies $\omega_\mu$) of qubits, of the interaction strength $\kappa_{ij}^\mu$ between qubits and the environment. An intuitive illustration is shown in Fig. \ref{R_3D}, where we use the blue dashed line to mark the rectification factor for $T_L=0.4\omega_0$ and $T_L=6\omega_0$, and the red dashed line to mark the case for $\omega_A=\omega_C=\omega$. Here we suppose qubits A and C are connected to separate but identical independent heat reservoirs and $\omega_B=5\omega_0$ and $\omega_C=2\omega-\omega_A$. In addition, we let $\omega$ take $\omega_0, 3\omega_0, 5\omega_0, 7\omega_0, 10\omega_0$, and $12\omega_0$ in Figs. \ref{R_3D}\hyperref[R_3D]{(a)}-\ref{R_3D}\hyperref[R_3D]{(f)} , respectively. We found that for $\omega<2\omega_B$ as shown in Figs. \ref{R_3D}\hyperref[R_3D]{(a)}-\ref{R_3D}\hyperref[R_3D]{(e)}, the greater the frequency difference of the two qubits connected with the left heat reservoir is, the better the rectification. For $\omega>2\omega_B$, one can find from Fig. \ref{R_3D}\hyperref[R_3D]{(f)} that $\omega_A=\omega_C=\omega$ corresponds to the best rectification effect. Comparing the red dashed lines in each picture, one can find that the larger difference in frequencies of qubits in contact with the left and right thermal reservoirs will lead to a better rectification effect. 

To further show the effect of the CHR on the rectification factor, let us focus on the green lines in Fig. \ref{R_3D}. As mentioned above, the steady state in the presence of the crossing dissipation depends on the initial state; nevertheless, the rectification factor doesn't since the heat current is proportional to the fraction $p$. Therefore, in Fig. \ref{R_3D}, we mainly address the steady state $\rho_2^{C,S}$ ( $p=1$) corresponding to the maximum heat current. It is shown in Figs. \ref{R_3D}\hyperref[R_3D]{(a)} and \ref{R_3D}\hyperref[R_3D]{(b)} that the green lines are under the dashed lines, which indicates that the common heat reservoir has an inhibitory effect on the rectification effect when $\omega<\omega_B$. But the green lines in Fig. \ref{R_3D}\hyperref[R_3D]{(c)} with $\omega\approx\omega_B$ show that CHR boosts the rectification effect. The boosting effect of CHR can be maintained even for $\omega_B<\omega<2\omega_B$ as shown in Figs. \ref{R_3D}\hyperref[R_3D]{(d)} and \ref{R_3D}\hyperref[R_3D]{(e)}. However, with the increasing of the frequency $\omega$, $\omega>2\omega_B$ as shown in Fig. \ref{R_3D}\hyperref[R_3D]{(f)}, the enhancement effect of CHR is only kept for nonequilibrium region, i.e., the temperature region except near $T_L=T_R$.
To sum up, when $\omega<2\omega_B$ in the absence of CHR, the rectification factor is large for $\omega_A$, greatly deviating from $\omega_C$. When $\omega>2\omega_B$ in the presence of CHR, $\omega_A=\omega_C$ will make a large rectification factor. However, the coupling strength $g_{\mu\nu}$ between qubits has no obvious effect on the rectification factor in the absence of CHR.

\begin{figure}
\begin{center}
		\includegraphics[width=8cm]{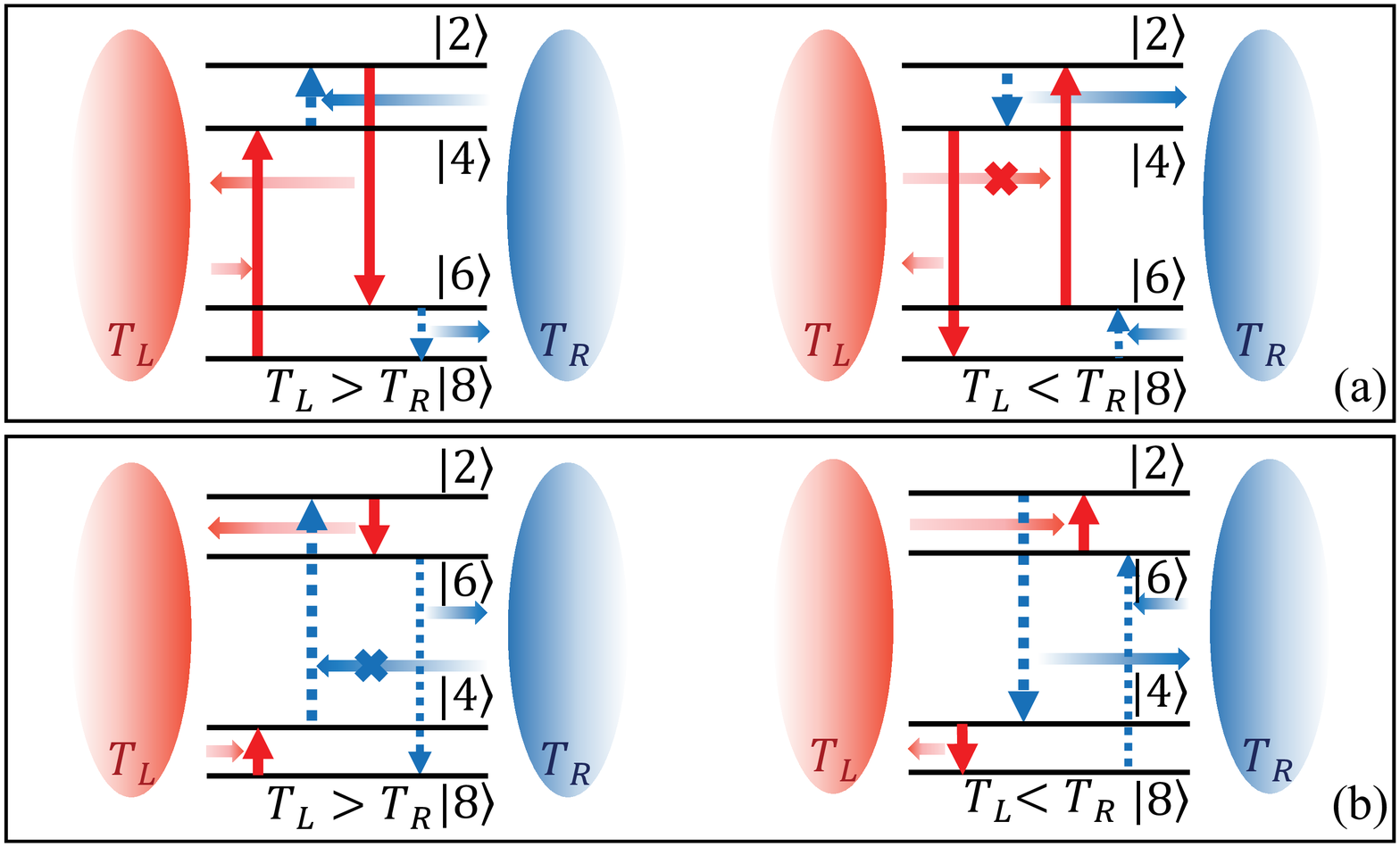}
\caption{A principle diagram for the quantum thermal diode in the presence of crossing dissipations. A simple transition cycle ($\vert8\rangle\rightarrow\vert4\rangle\rightarrow\vert2\rangle\rightarrow\vert6\rangle\rightarrow\vert8\rangle$) is taken as an example. (a) and (b) correspond to $\omega>\omega_B$ and $\omega<\omega_B$, respectively.}
\label{transition}
\end{center}
\end{figure}
To give an intuitive understanding of the quantum thermal diode, let's consider the transitions between the various eigenstates $\{\vert\tilde{i}\rangle\}$, which is illustrated in Fig. \ref{level}\hyperref[level]{(b)} implying four non-repeating transition cycles:
\begin{align}
\nonumber
\vert8\rangle&\rightarrow\vert4\rangle\rightarrow\vert2\rangle\rightarrow\vert6\rangle\rightarrow\vert8\rangle,\\
\nonumber
\vert8\rangle&\rightarrow\vert4\rangle\rightarrow\vert3\rangle\rightarrow\vert1\rangle\rightarrow\vert2\rangle\rightarrow\vert4\rangle\rightarrow\vert8\rangle,\\
\nonumber
\vert8\rangle&\rightarrow\vert4\rangle\rightarrow\vert3\rangle\rightarrow\vert1\rangle\rightarrow\vert2\rangle\rightarrow\vert6\rangle\rightarrow\vert8\rangle,\\
\vert8\rangle&\rightarrow\vert2\rangle\rightarrow\vert3\rangle\rightarrow\vert1\rangle\rightarrow\vert2\rangle\rightarrow\vert6\rangle\rightarrow\vert8\rangle.
\end{align} 
Fig. \ref{transition}\hyperref[transition]{(a)} corresponds to $\omega>\omega_B$, the transitions $\vert 8\rangle\leftrightarrow\vert 4\rangle$ and $\vert 6\rangle\leftrightarrow\vert 2\rangle$ induced by the left reservoir require the larger energy. When the temperature changes, the left reservoir will no longer induce the transition $\vert 6\rangle\leftrightarrow\vert 2\rangle$ easily, so this cycle cannot be completed and results in a difference in forward heat current and reverse one. When $\omega<\omega_B$ as shown in Fig. \ref{transition}\hyperref[transition]{(b)}, the transition $\vert 2\rangle\leftrightarrow\vert 4\rangle$ (or $\vert 5\rangle\leftrightarrow\vert 7\rangle$) and $\vert 6\rangle\leftrightarrow\vert 8\rangle$ induced by the right reservoir require a large amount of energy, and the transition cycle cannot be completed when the temperature $T_R$ is low. When the frequencies of qubits are close to each other, i.e., $\omega\approx\omega_B$, the transition rates of both reservoirs are nearly the same, which explains why the closer the qubits are to resonance, the worse the rectification effect is.

The thicker arrows in Fig. \ref{transition} indicate transitions between degenerate energy levels, which means there's a higher probability of transitions between these levels. Since only transition $\vert6\rangle\leftrightarrow\vert8\rangle$ induced by the right reservoir does not involve the degenerate energy level in this cycle, when $\omega_B$ is slightly larger than $\omega$, the inverse cycle corresponding to Fig. \ref{transition}\hyperref[transition]{(a)} is easier to complete than the positive cycle corresponding to Fig. \ref{transition}\hyperref[transition]{(b)}. Therefore, the system corresponding to Fig. \ref{transition}\hyperref[transition]{(b)} will have greater asymmetry, i.e., better rectification.
So the rectification at $\omega<\omega_B$ is slightly greater than that at $\omega>\omega_B$.

\section{Steady-state Mutual information}
\label{sec5}
In order to study the property of the steady state, we'd like to briefly introduce quantum entanglement, quantum discord (quantum correlation), classical correlation, and quantum mutual information. Quantum entanglement is usually considered as the most popular quantum correlation in a quantum system. For a bipartite ($L$ and $R$) system, a separable quantum state can be written as
\begin{align}
\rho_{LR}=\sum_kp_k\rho_L^k\otimes\rho_R^k,\quad\sum_kp_k=1,\quad p_k>0,\label{separable}
\end{align}
where $\rho_L^k$ and $\rho_R^k$ denote the local density matrices. If the quantum state cannot be written as Eq. (\ref{separable}), it is entangled \cite{hill1997entanglement,wootters1998entanglement,PhysRevA.64.032310,PhysRevA.85.062323,RevModPhys.81.865,hu2018steady,wang2019steady,chitambar2019quantum}.
 
However, quantum entanglement is not the unique quantum correlation in a quantum system. Quantum discord \cite{PhysRevLett.88.017901} is thought to be another type of quantum correlation different from quantum entanglement.
It is defined by the discrepancy of two classically equivalent quantum mutual information, or by the difference between the total correlation and the classical correlation. As we know, the total correlation is well described by the quantum mutual information 
\begin{align}
\mathcal{I}(\rho_{LR})=S(\rho_{L})+S(\rho_{R})-S(\rho_{LR}),
\end{align}
where $S(\rho)=-\mathrm{Tr}\rho\log\rho$ denotes the von Neumann entropy. The classical correlation is characterized by the most residual information after all potential measurements since measurements destroy quantum correlation. In this sense, the classical correlation is defined by
\begin{align}
\mathcal{C}(\rho_{LR})=S(\rho_L)-\inf\limits_{\{V_k^R\}}S(\rho_{LR}\vert\{V_k^R\}),\label{classical}
\end{align}
where $S(\rho_{LR}\vert\{V_k^R\})=\sum_k p_kS(\rho_{LR}\vert V_k^R)$ denotes the conditional entropy and the infimum is taken over all potential measurements $\{V_k^R\}$ performed on the subsystem $R$.
Thus the quantum discord is defined by
\begin{align}
\mathcal{Q}(\rho_{LR})=\mathcal{I}(\rho_{LR})-\mathcal{C}(\rho_{LR}).\label{nc}
\end{align}
It should be emphasized that quantum discord $\mathcal{Q}(\rho_{LR})$ is usually asymmetric under exchanging the subsystems $L$ and $R$. If we exchange $L$ and $R$, one will obtain the other discord denoted by $\mathcal{Q}(\rho_{RL})$. With these definitions, one can find that $\mathcal{Q}(\rho_{LR})=0$ for the quantum-classical states of the form $\rho_{LR}=\sum_k\rho^k_L\otimes\left\vert k\right\rangle_R\left\langle k\right\vert$, and $\mathcal{Q}(\rho_{RL})=0$ for the classical-quantum states of the form $\rho_{RL}=\sum_k\left\vert k\right\rangle_L\left\langle k\right\vert\otimes\rho^k_R$, where $ \{\left\vert k\right\rangle\}$ denotes the local basis of the corresponding subsytem. In particular, it is shown that the states of the form 
\begin{equation}\rho_{RL}=\sum_{k,k^\prime}\left\vert k\right\rangle_L\left\langle k\right\vert\otimes\left\vert k^\prime\right\rangle_R\left\langle k^\prime\right\vert
\end{equation} 
with $ \{\left\vert k'\right\rangle_R\}$ representing the basis of subsystem $R$ is classical-classical states which haven't any quantum correlation \cite{PhysRevA.77.042303,PhysRevLett.105.190502,PhysRevA.82.034302} with $\mathcal{Q}(\rho_{LR})=\mathcal{Q}(\rho_{RL})=0$. In this sense, one can obviously obtain $\mathcal{I}(\rho_{LR})=\mathcal{C}(\rho_{LR})$ for the states given in Eq. (\ref{nc}), namely, the total correlation of the states Eq. (\ref{nc}) is equal to the classical correlation.

Turning to our system, we regard atoms $A$ and $C$ in contact with the left common reservoir as subsystem $L$ and atom $B$ as subsystem $R$. so the composite system is $4\otimes 2$-dimensional. 
In the absence of the CHR, the steady state can be given as
\begin{align}
\nonumber
\rho^{I,S}&=(\rho_{11}^I\vert ++\rangle_L\langle ++\vert+\rho_{22}^I\vert +-\rangle_L\langle +-\vert\\
\nonumber
&+\rho_{55}^I\vert -+\rangle_L\langle -+\vert+\rho_{66}^I\vert --\rangle_L\langle --\vert)\otimes \vert +\rangle_R\langle +\vert\\
\nonumber
&+\rho_{33}^I\vert ++\rangle_L\langle ++\vert+\rho_{44}^I\vert +-\rangle_L\langle +-\vert\\
&+\rho_{77}^I\vert -+\rangle_L\langle -+\vert+\rho_{88}^I\vert --\rangle_L\langle --\vert)\otimes \vert -\rangle_R\langle -\vert,
\end{align}
and in the presence of CHR, the steady state Eq. (\ref{steady0}) can be given as
\begin{align}
\nonumber
\rho^{C,S}&=[(\rho^C_{22}+\rho^C_{25})\vert\psi^+\rangle_L\langle\psi^+\vert+(\rho^C_{22}-\rho^C_{25})\vert\psi^-\rangle_L\langle\psi^-\vert\\
\nonumber
&+\rho^C_{11}\vert ++\rangle_L\langle ++\vert+\rho^C_{66}\vert --\rangle_L\langle --\vert]\otimes \vert +\rangle_R\langle +\vert\\
\nonumber
&+[(\rho^C_{44}+\rho^C_{47})\vert\psi^+\rangle_L\langle\psi^+\vert+(\rho^C_{44}-\rho^C_{47})\vert\psi^-\rangle_L\langle\psi^-\vert\\
&+\rho^C_{33}\vert ++\rangle_L\langle ++\vert+\rho^C_{88}\vert --\rangle_L\langle --\vert]\otimes \vert -\rangle_R\langle -\vert,
\end{align}
where $\vert \psi^\pm\rangle_L=\frac{1}{\sqrt{2}}(\vert +-\rangle_L\pm\vert -+\rangle_L)$ represents the maximally entangled state of atoms A and C.

It is obvious that both $\rho^{I,S}$ and $\rho^{C,S}$ have the same form as Eq. (\ref{nc}), hence $\rho^{I,S}$ and $\rho^{C,S}$ are not only separable, but also classically-classically correlated. In particular, the classical correlation between $L$ and $R$ can be calculated based on Eq. (\ref{classical}) and Eq. (\ref{nc}) by the mutual information as
\begin{figure}
\begin{center}
		\includegraphics[width=8cm]{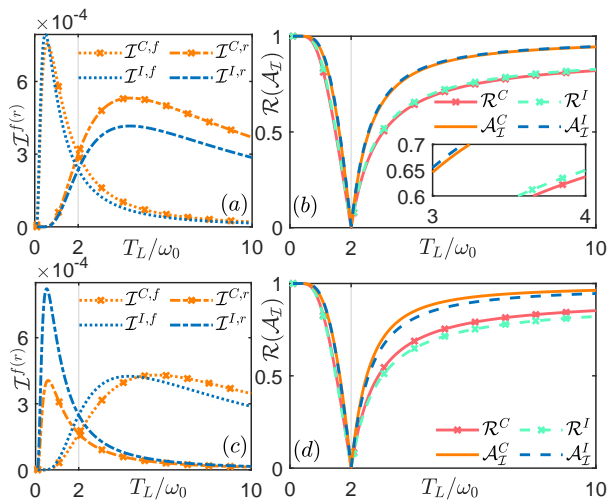}
\caption{Forward and reverse mutual information denoted by dotted lines and dashed-dotted lines, respectively, versus temperature in (a) and (c). Rectification factor $\mathcal{R}$ and asymmetric factor $\mathcal{A}_\mathcal{I}$ versus temperature $T_L$ in (b) and (d). Solid and dashed lines indicate the system contact with CHR or not. The inset in (b) shows an enlarged view at $T_L/\omega_0\in[3,4]$. Here, $\omega_0=1$, $\kappa=0.001\omega_0$, $T_R=\omega_0$, $g=g_{AC}=0.1\omega_0$, $p=1$, $\omega_A=\omega_C=\omega=\omega_0$, $\omega_B=5\omega_0$ in (a-b) and $\omega_A=\omega_C=\omega=5\omega_0$, $\omega_B=1\omega_0$ in (c-d).}
\label{R_AI}
\end{center}
\end{figure}
\begin{align}
\nonumber
\mathcal{C}(\rho^{I,S})=\mathcal{I}(\rho^{I,S})=&-\sum_{i=1}^8(\rho_{ii}^I)\log(\rho_{ii}^I),\\
\nonumber
\mathcal{C}(\rho^{C,S})=\mathcal{I}(\rho^{C,S})=&-\sum_{i=1}^2[(1-p)\tilde{\rho}_{ii}^S]\log [(1-p)\tilde{\rho}_{ii}^S]\\
&-\sum_{i=3}^8(p\tilde{\rho}_{ii}^S)\log (p\tilde{\rho}_{ii}^S),
\end{align}
where $\rho_{ii}^I$ is determined by Eq. (\ref{xcv}) and $\tilde{\rho}_{ii}^S$ is given in Appendix \ref{AppendixB}.

To describe the symmetry of the classical correlation (or quantum mutual information) induced by exchanging the temperature of the two heat reservoirs, we, similar to the heat current $\dot{Q}_\alpha^{f/r}$, define an asymmetric factor of mutual information $\mathcal{A}_\mathcal{I}$ as
\begin{equation}
\mathcal{A}_\mathcal{I}=\left|\frac{\left|\mathcal{I}^f\right|-\left|\mathcal{I}^r\right|}{\mathrm{Max}[\mathcal{I}^f,\mathcal{I}^r]}\right|,\label{defination_AI}
\end{equation}
where the superscripts $f$ and $r$ denote the forward and reverse mutual information. In Fig. \ref{R_AI}, we depict the rectification factor $\mathcal{R}$ and $\mathcal{A}_{\mathcal{I}}$ as a function of temperature $T_L$. One can find that $\mathcal{A}_{\mathcal{I}}$ has a good consistency with the $\mathcal{R}$, and CHR has the same effect on both of them. When $\omega<\omega_B$, as shown in Fig. \ref{R_AI}\hyperref[R_AI]{(b)}, the common heat reservoir inhibits both $\mathcal{R}$ and $\mathcal{A}_{\mathcal{I}}$. On the contrary, when $\omega>\omega_B$, as shown in Fig. \ref{R_AI}\hyperref[R_AI]{(d)}, the common heat reservoir contributes to these two quantities.
 Nevertheless, it is shown in Figs. \ref{R_AI}\hyperref[R_AI]{(a)} and \ref{R_AI}\hyperref[R_AI]{(c)} that $\mathcal{I}^f=\mathcal{I}^r\neq 0$ for $T_L=T_R$, which is different from the heat currents.
 
The beautiful consistency can also be well understood based on the transition depicted in Fig. \ref{transition}. In fact, one can find that Fig. \ref{transition} illustrates the asymmetric transition cycles induced by exchanging the reservoirs, which mainly determines the asymmetric transition rates. Thus a rough view is that the asymmetric transitions will further result in asymmetric populations, which determine the mutual information. In particular, the steady-state density matrices don't have any quantum correlation, which greatly simplifies the mathematical form of the state and makes the transition asymmetry to be greatly preserved in the asymmetry of populations. 

Before the end, we give a general understanding of the pure classical correlation in the steady state. Under the energy representation, for the case of IHR where there are no degenerate energy levels, the evolution of the density matrix is divided into two independent groups, separately corresponding to the diagonal elements and the off-diagonal elements  \cite{PhysRevE.76.031115}. During the evolution, the off-diagonal elements decay to zero, so the steady state is diagonal, which only includes the classical correlation. In the case of CHR, there exist degenerate energy levels. Similar to the case of IHR, the density matrix will also evolve to a diagonal steady state except for its degenerate subspace. Even though there could exist off-diagonal elements in the subspace, one can always find a proper local unitary operation to eliminate the off-diagonal elements. As a result, one can arrive at a diagonal steady state, which has only classical correlations. Note that the local operation on the subsystem $L$ does not affect the classical correlation between $L$ and $R$.

\section{Conclusions and discussion}
\label{sec6}
In this paper, we study the effects of three pairwise coupled qubits systems on heat transport between two heat reservoirs with temperature gradients. We find that energy levels of the system will degenerate, and crossing dissipation $\mathcal{L}_{AC}(\rho)$ will occur when two qubits $A$ and $C$ with the same natural frequency are coupled to another qubit $C$ with the same strength, i.e., $\omega_A=\omega_C=\omega$ and $g_{AB}=g_{BC}=g$. We find that the crossing dissipation always increases the heat current, but the effect on the rectification greatly depends on the frequencies and the internal couplings of the qubits. In particular, we show that in the presence of crossing dissipations, the steady-state heat current depends on the initial states, which provide a potential to adjust heat current by initial states, but the rectification factor is independent of the initial state of the system. 
The inverse current is also present in different dissipative channels at different temperature ranges in the presence of the crossing dissipations. Finally, we find that neither quantum entanglement nor quantum discord is present in the steady state, but the mutual information, i.e., the pure classical correlation has consistent asymmetric behavior with the heat rectification factor, which shows the important role of the classical correlation in the system.

Finally, we'd like to present the practical scenario for its potential usefulness. Suppose we own identical qubits with only $\sigma_z$-type couplings allowed, how can we realize the rectification? It is obvious that the bipartite coupling with $H_{SI}=\Delta\sigma^z_L\sigma^z_R$ won't produce any rectification effect due to the symmetric structure \cite{PhysRevE.95.022128}. We also find that the chain-type coupling won't contribute any rectification effect either. In this sense, our model provides an alternative method to realize a rectifier in the scenario. In addition, although we have shown the consistent behavior between classical correlation and the rectification effect, it will still be an interesting question whether or what features of quantum states are closely related to various quantum thermodynamical behaviors.

\section*{Acknowledgements}
This work was supported by the National Natural Science Foundation of China under Grant No.12175029, No. 12011530014, and No.11775040.

\appendix
\begin{widetext}
\section{Eigenvalues, eigen-operators and eigenfrequencies}
\label{AppendixA}
Here we will present some complicated expressions mentioned in the main text.
To give the global master equation, one needs to calculate the eigenvalues of the system's Hamiltonian. Corresponding to the computational bases, we can give the eigenvalues as
\begin{align}
\nonumber
\lambda_1=\frac{1}{2}(+\omega_A+\omega_B+\omega_C+g_{AB}+g_{BC}+g_{AC}),\quad\lambda_2=\frac{1}{2}(+\omega_A+\omega_B-\omega_C+g_{AB}-g_{BC}-g_{AC}),\\
\nonumber
\lambda_3=\frac{1}{2}(+\omega_A-\omega_B+\omega_C-g_{AB}-g_{BC}+g_{AC}),\quad\lambda_4=\frac{1}{2}(+\omega_A-\omega_B-\omega_C-g_{AB}+g_{BC}-g_{AC}),\\
\nonumber
\lambda_5=\frac{1}{2}(-\omega_A+\omega_B+\omega_C-g_{AB}+g_{BC}-g_{AC}),\quad\lambda_6=\frac{1}{2}(-\omega_A+\omega_B-\omega_C-g_{AB}-g_{BC}+g_{AC}),\\
\lambda_7=\frac{1}{2}(-\omega_A-\omega_B+\omega_C+g_{AB}-g_{BC}-g_{AC}),\quad\lambda_8=\frac{1}{2}(-\omega_A-\omega_B-\omega_C+g_{AB}+g_{BC}+g_{AC}).\label{Eqlevel}
\end{align}

Thus we can further calculate the eigen-operators and eigenfrequencies for the master equation as follows. Let
the eigen-operators and eigenfrequencies, respectively, denote by $\{V^\mu_m, m=1,...,4\}$ and $\omega_{ij}^\mu$ for qubit $\mu$ and the subscript $m$ label the different eigenfrequencies. A simple calculation can give
\begin{alignat}{3}
\nonumber
V_1^A=V_{15}^A&=\vert5\rangle\langle1\vert,\quad\omega_{15}^A &=\omega_A+g_{AB}+g_{AC},\quad V_2^A=V_{26}^A&=\vert6\rangle\langle2\vert,&\quad&\omega_{26}^A =\omega_A+g_{AB}-g_{AC},\\
\nonumber
V_3^A=V_{37}^A&=\vert7\rangle\langle3\vert,\quad\omega_{37}^A &=\omega_A-g_{AB}+g_{AC},\quad V_4^A=V_{48}^A&=\vert8\rangle\langle4\vert,&\quad&\omega_{48}^A =\omega_A-g_{AB}-g_{AC},\\
\nonumber
V_1^B=V_{13}^B&=\vert3\rangle\langle1\vert,\quad\omega_{13}^B &=\omega_B+g_{AB}+g_{BC},\quad V_2^B=V_{24}^B&=\vert4\rangle\langle2\vert,&\quad&\omega_{24}^B =\omega_B+g_{AB}-g_{BC},\\
\nonumber 
V_3^B=V_{57}^B&=\vert7\rangle\langle5\vert,\quad\omega_{57}^B &=\omega_B-g_{AB}+g_{BC},\quad V_4^B=V_{68}^B&=\vert8\rangle\langle6\vert,&\quad&\omega_{68}^B =\omega_B-g_{AB}-g_{BC},\\
\nonumber
V_1^C=V_{12}^C&=\vert2\rangle\langle1\vert,\quad\omega_{12}^C &=\omega_C+g_{BC}+g_{AC},\quad V_2^C=V_{34}^C&=\vert6\rangle\langle5\vert,&\quad&\omega_{34}^C =\omega_C+g_{BC}-g_{AC},\\
V_3^C=V_{56}^C&=\vert4\rangle\langle3\vert,\quad\omega_{56}^C &=\omega_C-g_{BC}+g_{AC},\quad V_4^C=V_{78}^C&=\vert8\rangle\langle7\vert,&\quad&\omega_{78}^C =\omega_C-g_{BC}-g_{AC}.
\end{alignat}
Note that the eigen-operator is $V_{ij}^\mu$ corresponds to the positive eigenfrequency, i.e., $\omega_{ij}^\mu>0$, otherwise ${V_{ij}^\mu}^\dagger$ corresponds to the negative eigenfrequency.

\section{Analytical steady states}
\label{AppendixB}
For the case that the left heat reservoir is IHR, the coefficient matrix in the dynamic $\vert\dot{\rho}^I\rangle=\mathcal{M}^I\vert\rho^I\rangle$ as $\mathcal{M}^I=\mathcal{M}^I_A+\mathcal{M}^I_B+\mathcal{M}^I_C$, where
\begin{align}
\nonumber
\mathcal{M}_A^I&=\mathrm{M}_{15}^A\otimes m^+\otimes m^++\mathrm{M}_{26}^A\otimes m^+\otimes m^-+\mathrm{M}_{37}^A\otimes m^-\otimes m^++\mathrm{M}_{48}^A\otimes m^-\otimes m^-,\\
\nonumber
\mathcal{M}_B^I&= m^+\otimes\mathrm{M}_{13}^B\otimes m^++ m^+\otimes\mathrm{M}_{24}^B\otimes m^-+ m^-\otimes\mathrm{M}_{57}^B\otimes m^++ m^-\otimes\mathrm{M}_{68}^B\otimes m^-,\\
\mathcal{M}_C^I&= m^+\otimes m^+\otimes\mathrm{M}_{12}^C+ m^+\otimes m^-\otimes\mathrm{M}_{34}^C+ m^-\otimes m^+\otimes\mathrm{M}_{56}^C+ m^-\otimes m^-\otimes\mathrm{M}_{78}^C,
\end{align}
with $\mathrm{M}^\mu_{ij}=2\left(
\begin{matrix}
-J_{\mu}(-\omega_{ij}^\mu)&J_{\mu}(+\omega_{ij}^\mu)\\
J_{\mu}(-\omega_{ij}^\mu)&-J_{\mu}(+\omega_{ij}^\mu)
\end{matrix}
\right)$, $ m^+=\left(
\begin{matrix}
1&0\\
0&0
\end{matrix}
\right)$, and $ m^-=\left(
\begin{matrix}
0&0\\
0&1
\end{matrix}
\right)$.
The steady state can be expressed as $\rho^{I,S}=\sum_{i=1}^8\rho_{ii}^I\vert i\rangle\langle i\vert$, where $\rho_{ii}^I=\frac{\rho_i^I}{N^I}$ is the normalized non-zero matrix element in the steady state with the normalization coefficient $N^I=\sum_{i=1}^8\rho^I_i$, which is not given here because its lengthy form. 

When there is no connection between the left and right subsystems, i.e., $g_{AB}=g_{BC}=0$, 
there are $A_{15}^\pm=A_{37}^\pm$, $A_{26}^\pm=A_{48}^\pm$, $B_{13}^\pm=B_{24}^\pm=B_{57}^\pm=B_{68}^\pm$, $C_{12}^\pm=C_{34}^\pm$, and $C_{56}^\pm=C_{78}^\pm$, the non-zero matrix elements of steady state are
\begin{alignat}{2}
\nonumber
\rho_1^I\vert_{g=0}&=B_{13}^+[A_{26}^+C_{13}^+(A_{15}^++C_{56}^-)+A_{15}^+C_{56}^+(A_{26}^-+C_{12}^+))],&\quad\rho_2^I\vert_{g=0}&=B_{13}^+[A_{26}^+C_{56}^-(A_{15}^-+C_{12}^-)+A_{15}^+C_{12}^-(A_{26}^++C_{56}^+))],\\
\nonumber
\rho_3^I\vert_{g=0}&=B_{13}^-[A_{26}^+C_{13}^+(A_{15}^++C_{56}^-)+A_{15}^+C_{56}^+(A_{26}^-+C_{12}^+))],&\quad\rho_4^I\vert_{g=0}&=B_{13}^-[A_{26}^+C_{56}^-(A_{15}^-+C_{12}^-)+A_{15}^+C_{12}^-(A_{26}^++C_{56}^+))],\\
\nonumber
\rho_5^I\vert_{g=0}&=B_{13}^+[A_{15}^-C_{12}^+(A_{26}^++C_{56}^+)+A_{26}^-C_{56}^+(A_{15}^++C_{12}^-))],&\quad\rho_6^I\vert_{g=0}&=B_{13}^+[A_{26}^-C_{12}^-(A_{15}^++C_{56}^-)+A_{15}^-C_{56}^-(A_{26}^-+C_{12}^+))],\\
\rho_7^I\vert_{g=0}&=B_{13}^-[A_{15}^-C_{12}^+(A_{26}^++C_{56}^+)+A_{26}^-C_{56}^+(A_{15}^++C_{12}^-))],&\quad\rho_8^I\vert_{g=0}&=B_{13}^-[A_{26}^-C_{12}^-(A_{15}^++C_{56}^-)+A_{15}^-C_{56}^-(A_{26}^-+C_{12}^+))].
\end{alignat}
It's immediate to get $\dot{Q}_\alpha\vert_{g=0}=0$.

When $\omega_A=\omega_C$ and $g_{AB}=g_{BC}=g$, namely, qubits A and C connect the genuine CHR, the coefficient matrix $\mathcal{M}_\alpha$ are
\begin{equation} 
\mathcal{M}_L^C=
\begin{pmatrix}
-4L_{12}^- & 4L_{12}^+ & 0 & 0 & 0 & 0 & 4L_{12}^+ & 0\\
2L_{12}^- & -2L_{12}^+-2L_{26}^- & 0 & 0 & 2L_{26}^+ & 0 & -2L_{12}^+-2L_{26}^- & 0\\
0 & 0 &-4L_{34}^- & 4L_{34}^+ &  0 & 0 & 0 & 4L_{34}^+\\
0 & 0 & 2L_{34}^- & -2L_{34}^+-2L_{48}^- & 0 & 2L_{48}^+ & 0 & -2L_{34}^+-2L_{48}^-\\
0 & 4L_{26}^- & 0 & 0 & -4L_{26}^+ & 0 & 4L_{26}^- & 0\\
0 & 0 & 0 & 4L_{48}^- & 0& -4L_{48}^+  & 0 & 4L_{48}^-\\
2L_{12}^- & -2L_{12}^+-2L_{26}^- & 0 & 0 & 2L_{26}^+ & 0 & -2L_{12}^+-2L_{26}^- & 0\\
0 & 0 & 2L_{34}^- & -2L_{34}^+-2L_{48}^- & 0 & 2L_{48}^+ & 0 & -2L_{34}^+-2L_{48}^-\\
\end{pmatrix},
\end{equation}
and
\begin{align}
\mathcal{M}_R^C=m^+\otimes \mathrm{M}_{12}^R\otimes m^++m^+\otimes \mathrm{M}_{24}^R\otimes m^-+m^-\otimes m^-\otimes \mathrm{M}_{24}^R +m^-\otimes m^+\otimes \mathrm{M}_{68}^R,
\end{align}
there are only six linearly independent equations in the corresponding matrix equations.

The steady state of the system, which is dependent on the initial state, is a linear combination of two particular steady states, i.e., $\rho^{C,S}=(1-p)\rho_1^{C,S}+p\rho_2^{C,S}$. 
In $\rho_1^{C,S}$, there are 
\begin{align}
\nonumber
\tilde{\rho}_{11}^S=\frac{R_{24}^+}{R_{24}^++R_{24}^-},\quad\tilde{\rho}_{22}^S=\frac{R_{24}^-}{R_{24}^++R_{24}^-}.
\end{align}

The normalized matrix element of the second state $\rho_2^{C,S}$ are $\tilde{\rho}_{ii}^S=\frac{\tilde{\rho}_i^S}{\tilde{N}_2}, i=3,...,8$, where
\begin{align}
\nonumber
 \tilde{\rho}_3^S&=4 L_{35}^+ L_{46}^+ L_{57}^+ L_{68}^+ R_{34}^+ 
+4 L_{35}^+ L_{46}^- L_{57}^+ L_{68}^+ R_{56}^+ 
 + 4 L_{35}^+ L_{46}^- L_{57}^+ L_{68}^- R_{78}^+
+ 2 L_{35}^+ L_{46}^+ L_{57}^+ R_{34}^+ R_{78}^+
 + 2 L_{35}^+ L_{46}^- L_{57}^+ R_{56}^+ R_{78}^+ \\
\nonumber
&+2L_{35}^+L_{46}^+L_{68}^+ R_{34}^+ R_{78}^- 
+ 2 L_{35}^+ L_{46}^- L_{68}^+ R_{56}^+ R_{78}^- 
+ 2 L_{35}^+ L_{57}^+ L_{68}^+ R_{34}^+ R_{56}^+ 
 + 2 L_{35}^+ L_{57}^+ L_{68}^- R_{34}^+ R_{78}^+  
+ 2 L_{46}^+ L_{57}^+ L_{68}^+ R_{34}^+ R_{56}^- \\
\nonumber
&+ 2 L_{46}^+ L_{57}^- L_{68}^+ R_{34}^+ R_{78}^-
 + L_{35}^+ L_{57}^+ R_{34}^+ R_{56}^+ R_{78}^+
+  L_{35}^+ L_{68}^+ R_{34}^+ R_{56}^+ R_{78}^-
 + L_{46}^+ L_{57}^+ R_{34}^+ R_{56}^- R_{78}^+ 
+ L_{46}^+ L_{68}^+ R_{34}^+ R_{56}^- R_{78}^-, \\
 \nonumber
 \tilde{\rho}_4^S&=4 L_{35}^+ L_{46}^+ L_{57}^+ L_{68}^+ R_{34}^- 
+ 4 L_{35}^- L_{46}^+ L_{57}^+ L_{68}^+ R_{56}^- 
+ 4 L_{35}^- L_{46}^+ L_{57}^- L_{68}^+ R_{78}^-
+ 2 L_{35}^+ L_{46}^+ L_{57}^+ R_{34}^- R_{78}^+ 
+ 2 L_{35}^- L_{46}^+ L_{57}^+ R_{56}^- R_{78}^+    \\
\nonumber
&+ 2 L_{35}^+ L_{46}^+ L_{68}^+ R_{34}^- R_{78}^-
+ 2 L_{35}^- L_{46}^+ L_{68}^+ R_{56}^- R_{78}^-
+ 2 L_{35}^+ L_{57}^+ L_{68}^+ R_{34}^- R_{56}^+ 
+ 2 L_{35}^+ L_{57}^+ L_{68}^- R_{34}^- R_{78}^+
+ 2 L_{46}^+ L_{57}^+ L_{68}^+ R_{34}^- R_{56}^-    \\
\nonumber
&+ 2 L_{46}^+ L_{57}^- L_{68}^+ R_{34}^- R_{78}^-
+ L_{35}^+ L_{57}^+ R_{34}^- R_{56}^+ R_{78}^+
+ L_{35}^+ L_{68}^+ R_{34}^- R_{56}^+ R_{78}^- 
+ L_{46}^+ L_{57}^+ R_{34}^- R_{56}^- R_{78}^+
+ L_{46}^+ L_{68}^+ R_{34}^- R_{56}^- R_{78}^-,\\
 \nonumber
 \tilde{\rho}_5^S&= 4 L_{35}^- L_{46}^+ L_{57}^+ L_{68}^+ R_{34}^+ 
 + 4 L_{35}^- L_{46}^- L_{57}^+ L_{68}^+ R_{56}^+ 
 + 4 L_{35}^- L_{46}^- L_{57}^+ L_{68}^- R_{78}^+ 
+  2 L_{35}^- L_{46}^+ L_{68}^+ R_{34}^+ R_{78}^- 
 + 2 L_{35}^- L_{46}^- L_{68}^+ R_{56}^+ R_{78}^-    \\
\nonumber
&+ 2 L_{35}^- L_{46}^+ L_{57}^+ R_{34}^+ R_{78}^+
  +2 L_{35}^- L_{46}^- L_{57}^+ R_{56}^+ R_{78}^+
+  2 L_{35}^- L_{57}^+ L_{68}^+ R_{34}^+ R_{56}^+ 
 + 2 L_{35}^- L_{57}^+ L_{68}^- R_{34}^+ R_{78}^+
+  2 L_{46}^- L_{57}^+ L_{68}^+ R_{34}^- R_{56}^+    \\
\nonumber
&+ 2 L_{46}^- L_{57}^+ L_{68}^- R_{34}^- R_{78}^+
 + L_{35}^- L_{57}^+ R_{34}^+ R_{56}^+ R_{78}^+ 
 + L_{35}^- L_{68}^+ R_{34}^+ R_{56}^+ R_{78}^-
 + L_{46}^- L_{57}^+ R_{34}^- R_{56}^+ R_{78}^+ 
 +L_{46}^- L_{68}^+ R_{34}^- R_{56}^+ R_{78}^-, \\
 \nonumber
 \tilde{\rho}_6^S&=4 L_{35}^+ L_{46}^- L_{57}^+ L_{68}^+ R_{34}^- 
 + 4 L_{35}^- L_{46}^- L_{57}^+ L_{68}^+ R_{56}^- 
 + 4 L_{35}^- L_{46}^- L_{57}^- L_{68}^+ R_{78}^-
 + 2 L_{35}^+ L_{46}^- L_{57}^+ R_{34}^- R_{78}^+
 + 2 L_{35}^- L_{46}^- L_{57}^+ R_{56}^- R_{78}^+    \\
\nonumber
&+ 2 L_{35}^+ L_{46}^- L_{68}^+ R_{34}^- R_{78}^-
 + 2 L_{35}^- L_{46}^- L_{68}^+ R_{56}^- R_{78}^- 
 + 2 L_{35}^- L_{57}^+ L_{68}^+ R_{34}^+ R_{56}^-  
 + 2 L_{35}^- L_{57}^- L_{68}^+ R_{34}^+ R_{78}^- 
 + 2 L_{46}^- L_{57}^+ L_{68}^+ R_{34}^- R_{56}^-     \\
\nonumber
&+ 2 L_{46}^- L_{57}^- L_{68}^+ R_{34}^- R_{78}^- 
 + L_{35}^- L_{57}^+ R_{34}^+ R_{56}^- R_{78}^+
 +L_{35}^- L_{68}^+ R_{34}^+ R_{56}^- R_{78}^-  
 + L_{46}^- L_{57}^+ R_{34}^- R_{56}^- R_{78}^+
 + L_{46}^- L_{68}^+ R_{34}^- R_{56}^- R_{78}^-,\\
 \nonumber
 \tilde{\rho}_7^S&=4 L_{35}^- L_{46}^+ L_{57}^- L_{68}^+ R_{34}^+ 
 + 4 L_{35}^- L_{46}^- L_{57}^- L_{68}^+ R_{56}^+ 
 + 4 L_{35}^- L_{46}^- L_{57}^- L_{68}^- R_{78}^+
 + 2 L_{35}^- L_{46}^+ L_{57}^- R_{34}^+ R_{78}^+ 
 + 2 L_{35}^- L_{46}^- L_{57}^- R_{56}^+ R_{78}^+     \\
\nonumber
&+ 2 L_{35}^+ L_{46}^- L_{68}^- R_{34}^- R_{78}^+ 
 + 2 L_{35}^- L_{46}^- L_{68}^- R_{56}^- R_{78}^+
 + 2 L_{35}^- L_{57}^- L_{68}^+ R_{34}^+ R_{56}^+ 
 + 2 L_{35}^- L_{57}^- L_{68}^- R_{34}^+ R_{78}^+ 
 + 2 L_{46}^- L_{57}^- L_{68}^+ R_{34}^- R_{56}^+     \\
\nonumber
&+ 2 L_{46}^- L_{57}^- L_{68}^- R_{34}^- R_{78}^+ 
 + L_{35}^- L_{57}^- R_{34}^+ R_{56}^+ R_{78}^+
 + L_{35}^- L_{68}^- R_{34}^+ R_{56}^- R_{78}^+  
 + L_{46}^- L_{57}^- R_{34}^- R_{56}^+ R_{78}^+
 + L_{46}^- L_{68}^- R_{34}^- R_{56}^- R_{78}^+,\\
 \nonumber
 \tilde{\rho}_8^S&=4 L_{35}^+ L_{46}^- L_{57}^+ L_{68}^- R_{34}^- 
 + 4 L_{35}^- L_{46}^- L_{57}^+ L_{68}^- R_{56}^- 
 + 4 L_{35}^- L_{46}^- L_{57}^- L_{68}^- R_{78}^-
 + 2 L_{35}^- L_{46}^+ L_{57}^- R_{34}^+ R_{78}^- 
 + 2 L_{35}^- L_{46}^- L_{57}^- R_{56}^+ R_{78}^-    \\
\nonumber
&+ 2 L_{35}^+ L_{46}^- L_{68}^- R_{34}^- R_{78}^-
 + 2 L_{35}^- L_{46}^- L_{68}^- R_{56}^- R_{78}^-
 + 2 L_{35}^- L_{57}^+ L_{68}^- R_{34}^+ R_{56}^- 
 + 2 L_{35}^- L_{57}^- L_{68}^- R_{34}^+ R_{78}^-  
 + 2 L_{46}^- L_{57}^+ L_{68}^- R_{34}^- R_{56}^-     \\
&+ 2 L_{46}^- L_{57}^- L_{68}^- R_{34}^- R_{78}^- 
 + L_{35}^- L_{57}^- R_{34}^+ R_{56}^+ R_{78}^-
 + L_{35}^- L_{68}^- R_{34}^+ R_{56}^- R_{78}^- 
 + L_{46}^- L_{57}^- R_{34}^- R_{56}^+ R_{78}^- 
 +L_{46}^- L_{68}^- R_{34}^- R_{56}^- R_{78}^- ,
\end{align}
where $\alpha^\pm_{ij}$ is the spectral density $J_\alpha(\pm\omega^\alpha_{ij})$, and the normalization coefficient $\tilde{N}_2=\tilde{\rho}_3^S+\tilde{\rho}_4^S+\tilde{\rho}_5^S+\tilde{\rho}_6^S+\tilde{\rho}_7^S+\tilde{\rho}_8^S$.

It's intuitively easy to understand that the heat current is zero when $g=0$ since $g$ describes the conduction of the left and right subsystems. In terms of analytic expressions, there is $\tilde{\omega}_{35}^L=\tilde{\omega}_{46}^L$, $\tilde{\omega}_{57}^L=\tilde{\omega}_{68}^L$, and $\tilde{\omega}_{34}^R=\tilde{\omega}_{12\&56}^R=\tilde{\omega}_{78}^R$ when $g=0$, the unnormalized matrix elements can be expressed as
\begin{alignat}{2}
\nonumber
\tilde{\rho}_3^S\vert_{g=0}&=L_{35}^+L_{57}^+R_{34}^+,\quad \tilde{\rho}_4^S\vert_{g=0}&=L_{35}^+L_{57}^+R_{34}^-,\\
\nonumber
\tilde{\rho}_5^S\vert_{g=0}&=L_{35}^-L_{57}^+R_{34}^+,\quad \tilde{\rho}_6^S\vert_{g=0}&=L_{35}^-L_{57}^+R_{34}^-,\\
\tilde{\rho}_7^S\vert_{g=0}&=L_{35}^-L_{57}^-R_{34}^+,\quad \tilde{\rho}_8^S\vert_{g=0}&=L_{35}^-L_{57}^-R_{34}^-.
\end{alignat}
When we substitute the above equation into the heat current Eq. (\ref{steady_Q}), it's easy to get $\dot{Q}_\alpha\vert_{\tilde{\rho}^S,g=0}=0$ since $\dot{Q}_\alpha\vert_{\tilde{\rho}_1^S}=0$ and $\dot{Q}_\alpha\vert_{\tilde{\rho}^S_2,g=0}=0$.
\end{widetext}

\section{Analytical steady-state heat currents}
\label{AppendixC}
When the left reservoir is IHR, the heat transport between the system composed of three atoms and the environment can be expressed analytically as
\begin{align}
\nonumber
\dot{Q}_A\vert_{\rho^{I,S}}=&-(\omega_{15}^A\Gamma_{15}^A+\omega_{26}^A\Gamma_{26}^A+\omega_{37}^A\Gamma_{37}^A+\omega_{48}^A\Gamma_{48}^A),\\
\nonumber
\dot{Q}_B\vert_{\rho^{I,S}}=&-(\omega_{13}^B\Gamma_{13}^B+\omega_{24}^B\Gamma_{24}^B+\omega_{57}^B\Gamma_{57}^B+\omega_{68}^B\Gamma_{68}^B),\\
\dot{Q}_C\vert_{\rho^{I,S}}=&-(\omega_{12}^C\Gamma_{12}^C+\omega_{56}^C\Gamma_{56}^C+\omega_{34}^C\Gamma_{34}^C+\omega_{78}^C\Gamma_{78}^C).\label{QABC}
\end{align}
When the left reservoir is CHR, the heat current from the heat reservoir $B_\alpha$ to the system can be written as
\begin{align}
\nonumber
\dot{Q}_L\vert_{\rho_2^{C,S}}=&-2(\omega_{12}^L\Gamma_{12}^L+\omega_{26}^L\Gamma_{26}^L+\omega_{34}^L\Gamma_{34}^L+\omega_{48}^L\Gamma_{48}^L)\\
\nonumber
&-4[\omega_{26}^LJ_L(-\omega_{26}^L)-\omega_{12}^LJ_L(+\omega_{12}^L)]\tilde{\rho}_{55}^S\\
\nonumber
&-4[\omega_{48}^LJ_L(-\omega_{48}^L)-\omega_{34}^LJ_L(+\omega_{34}^L)]\tilde{\rho}_{66}^S,\\
\dot{Q}_R\vert_{\rho_2^{C,S}}=&-(\omega_{13}^R\Gamma_{13}^R+2\omega_{24}^R\Gamma_{24}^R+\omega_{68}^R\Gamma_{68}^R).\label{steady_Q}
\end{align}

When crossing dissipation is considered, the system has two independent steady states. The state $\rho_1^{C,S}$ blocks heat current, and the heat currents for this state in the independent and the crossing dissipation channel are respectively
\begin{align}
\nonumber
\dot{Q}_L^d\vert_{\rho_1^{C,S}}=&4\{\tilde{\rho}_{11}^S[\omega_{12}^LJ_L(+\omega_{12}^L)-\omega_{26}^LJ_L(-\omega_{26}^L)]\\
\nonumber
&+\tilde{\rho}_{22}^S[\omega_{34}^LJ_L(+\omega_{34}^L)-\omega_{48}^LJ_L(-\omega_{48}^L)]\},\\
\dot{Q}_L^c\vert_{\rho_1^{C,S}}=&-\dot{Q}_L^d\vert_{\rho_1^{C,S}}.
\end{align}
The steady state $\rho^{C,S}_2$ allows the maximum heat current, and the heat current for this state in the two channels can be expressed as
\begin{align}
\nonumber
\dot{Q}_L^d\vert_{\rho^{C,S}_2}=&2(\omega_{12}^L\Gamma_{12}^L+\omega_{26}^L\Gamma_{26}^L+\omega_{34}^L\Gamma_{34}^L+\omega_{48}^L\Gamma_{48}^L),\\
\nonumber
\dot{Q}_L^c\vert_{\rho^{C,S}_2}=&4\{[\omega_{12}^LJ_L(+\omega_{12}^L)-\omega_{26}^LJ_L(-\omega_{26}^L)]\tilde{\rho}_{55}^S\\
&+[\omega_{34}^LJ_L(+\omega_{34}^L)-\omega_{48}^LJ_L(-\omega_{48}^L)]\tilde{\rho}_{66}^S\}.\label{Qcd2}
\end{align}
When $T_L=0$, there is $n_L(\omega^L_{ij})=0$ for any eigenfrequency $\omega_{ij}^L$. So the system will evolve to the state $\rho_2^{C,S}=\tilde{\rho}^S_{77}\vert\tilde{7}\rangle\langle\tilde{7}\vert+\tilde{\rho}^S_{88}\vert\tilde{8}\rangle\langle\tilde{8}\vert$. In this state, the heat currents in two channels vanish, that is
\begin{align}
\dot{Q}_L^d\vert_{\rho_2^{C,S},T_L=0}=-\dot{Q}_L^c\vert_{\rho_2^{C,S},T_L=0}=0.\label{Qcd2TL}
\end{align}
When $T_R=0$, we have $n_R(\omega^R_{ij})=0$ and $\tilde{\rho}_{11}^S=\tilde{\rho}_{33}^S=\tilde{\rho}_{44}^S=\tilde{\rho}_{77}^S=0$, so Eq. (\ref{Qcd2}) can be simplified as
\begin{align}
\dot{Q}_L^d\vert_{\rho_2^{C,S},T_R=0}=-\dot{Q}_L^c\vert_{\rho_2^{C,S},T_R=0}=4\kappa(\omega_{26}^L\tilde{\rho}_{55}^S+\omega_{48}^L\tilde{\rho}_{66}^S).\label{Qcd2TR}
\end{align}
So the total heat current of the system also disappears for $T_R=0$.

\begin{figure}
\begin{center}
		\includegraphics[width=8cm]{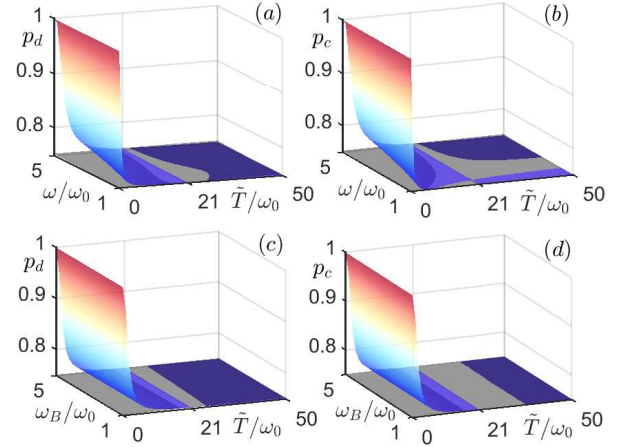}
\caption{$p_d$ and $p_c$ versus the atomic frequency $\omega$ (and $\omega_B$) and the temperature $\tilde{T}$ in (a-b) (and (c-d)). Colored (upper) and shaded (lower) surfaces correspond to $\tilde{T}=T_L$ with $T_R=21\omega_0$, and $\tilde{T}=T_R$ with $T_L=21\omega_0$, respectively. Here, $\omega_0=1$, $g=g_{AC}=0.1\omega_0$, $\kappa=0.001\omega_0$, $\omega_B=5\omega_0$ in (a-b), and $\omega=3\omega_0$ in (c-d).}
\label{pcpd_omega}
\end{center}
\end{figure}

\section{The dependence of $p_{d(c)}$ on the parameters of the system}
\label{AppendixD}
\begin{figure}
\begin{center}
		\includegraphics[width=8cm]{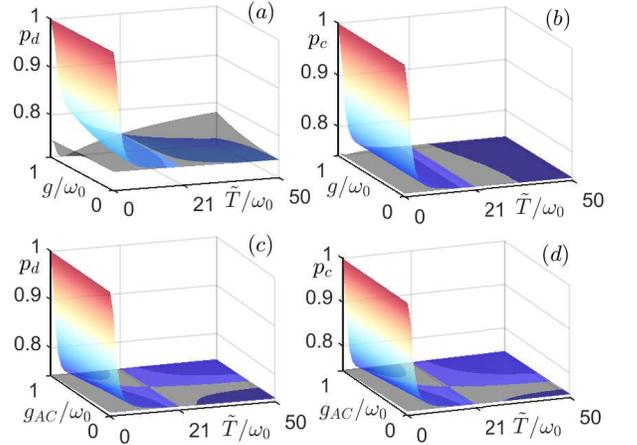}
\caption{$p_d$ and $p_c$ versus the atomic coupling strength $g$ (and $g_{AC}$) and the temperature $\tilde{T}$ in (a-b) (and (c-d)). $g_{AC}=0.1\omega_0$ in (a-b), and $g=0.1\omega_0$ in (c-d). What the colored (upper) and shaded (lower)  surfaces denote and all the other taken parameters are the same as Fig. \ref{pcpd_omega}.}
\label{pcpd_g}
\end{center}
\end{figure}
We have plotted $p_d$ (or $p_c$) versus varous parameters of the system in Fig. \ref{pcpd_omega} and Fig. \ref{pcpd_g}. It is shown that the frequencies of either the atoms $A$ and $C$ connected to CHR or the atom $B$ connected to IHR have almost no effect on $p_c$ and $p_d$. Figs. \ref{pcpd_g}\hyperref[pcpd_g]{(a)} and \ref{pcpd_g}\hyperref[pcpd_g]{(b)} indicate that $g$ has a small effect on $p_d$ and a negligible effect on $p_c$. When $\tilde{T}$ has a relatively large value, $p_d$ (or $p_c$) tends to a constant, which shows negligible effect of whether exchanging the hot and the cold terminals or not. However, if $\tilde{T}$ is small, exchanging the two terminals will indicate significant effect on $p_d$ and $p_c$.

\bibliography{CHR_diode}

\end{document}